\documentclass[12pt]{article}
\usepackage{fullpage, amsthm, amsmath}
\usepackage{mathtools}
\usepackage{graphicx, psfrag, amsfonts, bm, epsfig}
\usepackage{natbib}
\usepackage{subfigure}
\usepackage{hyperref}

\usepackage{graphicx, psfrag, amsfonts, setspace,lscape,longtable}
\usepackage{amssymb}
\usepackage{mathrsfs}
\usepackage{url}
\newcommand*{\rom}[1]{\expandafter\@slowromancap\romannumeral #1@}

\newcommand{\ba}{\bm a}
\newcommand{\bmm}{\bm m}
\newcommand{\bt}{\bm t}

\newcommand{\bC}{\bm{C}}

\newcommand{\eb}{\bm{e}}
\newcommand{\bb}{\bm{b}}

\newcommand{\bw}{\bm{w}}

\newcommand{\bX}{\bm{X}}

\newcommand{\bO}{\bm{0}}

\newcommand{\bmu}{\bm{\mu}}

\newcommand{\bbeta}{\bm{\beta}}

\newcommand{\cT}{{\cal T}}

\newtheorem{theorem}{Theorem}
\newtheorem{lemma}{Lemma}
\newtheorem{definition}{Definition}

\newcommand{\is}{\itemsep=0pt}
\newcommand{\bd}[1]{\begin{description}[#1]\is}
  \newcommand{\ed}{\end{description}}
\newcommand{\bi}{\begin{itemize}\is}
  \newcommand{\ei}{\end{itemize}}
\newcommand{\be}{\begin{enumerate}\is}
  \newcommand{\ee}{\end{enumerate}}

\usepackage[dvipsnames,usenames]{color}

\begin{document}
\doublespacing

\title{A Latent Gaussian Process Model with Application to Monitoring Clinical Trials}

\author{
           Yanxun Xu   \\
           {\small Division of Statistics and Scientific Computing, The University of Texas at Austin, Austin, TX}
           \and
           Yuan Ji \thanks{Address for correspondence: NorthShore University HealthSystem / The
     University of Chicago, 1001 University Place, Evanston, Illinois, USA 60091. Email: jiyuan@uchicago.edu}\\
          {\small   Center for Biomedical Informatics, NorthShore University HealthSystem, Evanston, IL}\\
          {\small Department of Health Studies, The University of Chicago, Chicago, IL}          
}
\date{}
\maketitle

\begin{abstract}
In many clinical trials treatments need to be repeatedly applied
as diseases relapse frequently after remission over a long period
of time (e.g., 35 weeks). Most research in
statistics focuses on the overall trial design, such as sample size
and power calculation, or on the data analysis after trials are
completed. Little is done to improve the efficiency of trial
monitoring, such as early termination of trials due to futility. The
challenge faced in such trial monitoring is mostly caused by the
need to properly model repeated outcomes from patients. We propose a
Bayesian trial monitoring scheme for clinical trials with repeated and
potentially cyclic binary outcomes. We construct a latent Gaussian process (LGP) to model discrete
longitudinal data in those trials.  
LGP describes the underlying latent process that gives rise to the
observed longitudinal binary outcomes. The posterior consistency property of the proposed model is studied. Posterior inference is conducted with a hybrid Monte Carlo algorithm. Simulation studies are conducted under various clinical scenarios, and a case study is reported based on a real-life
 trial. Matlab program for implementing the interim monitoring procedure is freely
 available at 
 \href{http://www.ma.utexas.edu/users/yxu}{{\color{blue}http://www.ma.utexas.edu/users/yxu}}. 
 
\end{abstract}

\section{Introduction}
\label{sec:Intro}
\subsection{Background}

We consider Bayesian monitoring for clinical trials with longitudinal
binary outcomes, such as multiple disease remissions over time. This
is an area that has received little attention in the trial design
field. Most work has focused on the overall design of the trial
\citep{frison1992repeated, raudenbush2001effects, galbraith2002guidelines, hedeker1999sample}, such
as sample size calculation and choice of followup time points, or
methods for analyzing data after trials are completed \citep{liang1986longitudinal,hedeker2006longitudinal}. In many
trials, the primary outcome data for each subject are recurrent and cyclic based on a longitudinal binary variable.  For example, a data vector of
interest could be binary values observed on consecutive time points,
and the vector values alternate between 1's and 0's. This type
of data arises from clinical trials in many therapeutic areas, such as
auto-immune diseases like multiple sclerosis or
Schizophrenia (see \citealp{goldman2010possible} and references
therein). Typically, treatments need to be applied multiple times
in hopes 
to slow down disease progression. When multiple drugs are to be
evaluated, it is challenging to quantify and compare the
efficacy of the drugs 
due to the temporal and periodic nature of disease progression and
treatment response. For example, when two drugs are compared in a
clinical trial involving one new treatment and a standard treatment,
therapeutic effects are usually assessed by comparing the relapse
rates of treatments within a time framework. 
 However, that strategy 
might not be ideal since a treatment with a slightly inflated relapse rate
might still be preferred if it allows patients to stay in disease
remission for a longer period of time. 

In this paper, we consider an important design issue for trials of the
same nature.  For each patient, multiple responses to treatments and
multiple disease relapses are expected. One goal is to demonstrate that one treatment allows patients to stay in disease remission longer than the other, and a second goal is to predict future responses for individual patients based on observed outcomes. 
A typical
patient response vector may look like $(1, 1, 0, 0, 1, 1, 1, 0,0
\cdots) $ with binary outcomes at multiple time points, where
``1'' represents a response and ``0'' a nonresponse.  In addition,
each binary outcome is associated with a time point at which the
outcome is recorded. Therefore, the data is summarized as $\{e(t_k),
\; k=1, \ldots, K\}$ where $e(t_k) \in \{0,1\}$ is binary. In  \cite{comi2012placebo}, a response $\{ e(t_k) = 1\}$ is defined as a combination of cognitive
improvements assessed by psychiatric tests. 
 An important presumption that underpins 
 this type of trial is that the
 unobserved  disease progression is a continuous process over time,
 although measurements of outcome can only be taken at
 discrete time points, e.g., once a month.  In other words, the
 observed binary longitudinal outcomes are manifestations of latent
 continuous processes. 

To model a latent process, \citet{zeger1985analysis} considered an extension of logistic regression to binary longitudinal observations. However, their method was only applied to stationary binary series. 
\cite{czado2008state} developed a state space mixed model for  binary
longitudinal observations using the standard linear state-space
formulation \citep{kalman1963mathematical}. But the linear
state-space models are limited and may not work well on the
aforementioned trials with nonlinear and cyclic
responses. \cite{hall2008modelling} proposed a latent Gaussian process
model for sparse generalized longitudinal observations. They estimated
the mean function by smoothing the mean of all trajectories and
covariance kernel by calculating functional principle components as
the eigenfunctions. While useful, their smoothing technique must be
applied to all observations collectively, not allowing random
effects that deviate from the population curve. More importantly, available  models for longitudinal data cannot
accommodate all our needs for trial monitoring, 
which include 1) to model cyclic binary responses, 2) to forecast
future outcomes for each patient based on the cyclic pattern, 3)
to compare population patterns between different conditions (treatment
v.s. control), 4) to reflect the underlying continuous disease
progression mechanism. Also, to our knowledge, there does not exist an
adaptive monitoring scheme for trials with binary longitudinal 
outcomes. For example, there are no existing statistical methods for
terminating such trials when the treatment arm is not effective as the
control. To this end, 
we consider a latent stochastic process with cyclic
 features to describe the relapsing nature of the disease. Specifically,
 let  $a_{j}(t)=\mu(t)+\tau_j(t)$ be a sum of the mean process
 $\mu(t)$
 describing the treatment effect over time, and a cyclic process
 $\tau_j(t)$ describing the subject-$j$-specific recurrent disease
 progression. For example, the treatment effect $\mu(t)$ could
 be increasing over time due to the continuous usage of the drug but
 $\tau_j(t)$, to be modeled as the latent
 Gaussian process, could be
 cyclic mimicking the recurrent disease progression as a result of the
 battle between the disease-causing antigens and the disease-fighting
 immune cells. We model the binary outcome at a time point $t_k$
 as an indicator $e_j(t_k) = I\{a_j(t_k)>a_h\}$ with a fixed and
 arbitrary threshold value
 $a_h$. This construction can be considered a stochastic-process
 version of the latent probit model in \citet{Albert-Chib-1993}. We
 defer the construction of $a_j(t)$ to Section \ref{sec:model}.  

\subsection{Gaussian Process}

Gaussian process (GP)  has been frequently applied to research areas in finance, engineering, cognitive
research,  and others.  Examples of such applications include
work in machine learning \citep{Rasmussen:2006}, neural
networks \citep{Neal1995}, and batched data \citep{Shi2007}.  Initially introduced in \citet{Hagan1978},  GP priors have also been
used in Bayesian inference for regression and classification problems;
see a review by  \citet{Williams1998}.  GP models are considered
nonparametric in curve fitting as they are not dependent on any functions to describe the shapes of the curves. At the same
time, GP models are relatively easy to compute since they are based on
multivariate Gaussian distributions. 

A Gaussian process is a stochastic process $\ba(t),$ 
for which any $n$-finite variates $\ba_n = \{a(t_1), \ldots,
a(t_n)\}^\prime$ has a multivariate Gaussian distribution given by  
$$P(\ba_n \mid \bmu, \bC) \propto |\bC|^{-\frac{1}{2}} \exp \left\{-\frac{1}{2}(\ba_n-\bmu)'\bC^{-1}(\ba_n-\bmu)\right\}$$ 
for any $n$ and collection of input $\{t_1, \dots, t_n\}$. Vector
$\bmu$ is the mean vector with dimension $n$ 
and $\bC$ represents the $n\times n$
covariance matrix, and is often parameterized as a 
covariance function $C^{uv}(t_u, t_v; \Theta)$, where $\Theta$ is a
set of hyperparameters. In other words, a Gaussian process is one
which 
every finite-dimensional joint distribution is multivariate
Gaussian. 
Since we are expressing the correlations between different points in
the input space through the covariance function, it is crucial to
study the characteristics of the GP, such as its smooth properties
and differentiability \citep{abrahamsen1997review}.

\subsection{Motivating Examples}
The key idea of our proposed LGP is to
model observed discrete longitudinal outcomes 
by thresholding the latent variates that follow a GP prior. 
We consider two motivating examples. The first example is a randomized, 
double-blind, placebo-controlled clinical trial aiming  to
evaluate the efficacy and safety of a 200-mcg dose of a new drug
(drug name masked for copyright protection) 
  in
patients with systemic lupus erythematosus (SLE).  We will refer to
this trial as the ``lupus trial" hereinafter. SLE  is a
long-term auto-immune disease in which 
 the body 
 mistakenly attacks healthy organs, such as the skin and brain. 
 There is no cure for the disease, and immuno-therapies only reduce symptoms and must be applied
frequently.  Patients respond to treatments quickly, typically within weeks, although the
 disease also relapses quickly. A clinical response is based on the SLE
 responder index, which is defined by a combination of four different
 cognitive test scores.
In the presence of control, each patient is randomized and
followed for 35 weeks, during which time multiple relapses and
responses might be observed. The trial objective is to compare the response
rates 
of the new treatment and the control. It is desirable to terminate the trial early since the
35-week follow-up time period is long; for example, whenever there is substantial
evidence that the new treatment is no better than the control, the
trial should be stopped for both ethical and financial considerations.
 The total sample size for the lupus trial is 200 patients,  who are randomized 1:1
 between
 the two treatments. 
 
A second similar example is a placebo-controlled trial of oral laquinimod for multiple sclerosis \citep{comi2012placebo}. 
The efficacy and safety of laquinimod in patients with
relapsing-remitting multiple sclerosis were evaluated in a randomized,
double-blind, phase III trial with 1,106 patients randomly
assigned in a 1:1 ratio to receive placebo or laquinimod. The primary
end point was whether or not the disease had relapsed during the study
period. An event was counted as a relapse if the patient's symptoms
were accompanied by objective neurologic changes according to
predefined criteria.

A common feature in both trials is that multiple responses and relapses are expected for each patient
during the follow-up period, and therefore simple monotonic parametric dose-response models
such as logistic regression are no longer suitable. The observed data
for each patient will be a binary vector ${\bm e}=(e_1,e_2,
\dots,e_K)$ for $K$ time points. The indicators are 
temporally correlated.

In Section \ref{sec:model}, we present probability models and computational methods
based on the posterior distributions. In Section \ref{sec:post}, we derive 
posterior consistency results. In Section \ref{sec:stop}, we propose inference for
predicting future responses and introduce stopping rules as part of
the trial design. We examine the performance of LGP through extensive
simulation studies in Section \ref{simulation}. In Section
\ref{sec:trial}, we report numerical results based on the lupus
trial.  Finally, we conclude with a brief
discussion in Section \ref{sec:dis}. 


\section{Probability Model}
\label{sec:model}
\subsection{Latent Gaussian Process }
\label{probabilitymodel}
In the aforementioned lupus trial, patients are
randomized between two arms,  with arm  1 being the standard control and
arm 2 the new treatment. Multiple interim analyses are proposed to
monitor the progress of the trial and to compare the two treatments. 
At the time of interim analysis, assume that $N_1$ and $N_2$ patients have been assigned to
 arms 1 and 2, respectively ($N_1=N_2$ if equal randomization). For
 each patient $j$ in group $i$, we assume that disease outcomes have
 been measured at  $K_{ij}$  time points, denoted as $\bt_{ij}=(t_1,
 t_2, \dots,t_{K_{ij}})'$. Here a time point refers to the
 duration of follow-up. Let $e_{ij}(t_k)$ be the binary response at time point $t_k$, simplified as $e_{ijk}$. If the
 $j$th patient in group $i$ responded at time $t_k$, then $e_{ijk}=1$;
 otherwise, $e_{ijk}=0$. Therefore, the observed data are tuple
 $\eb^K=\{e_{ijk}\}, i=1,2; ~j=1,2, \dots, N_i; ~k=1, 2, \dots,
 K_{ij}$.
 
Define a stochastic process 
\begin{equation}
a_{ij}(t)=\mu_{i}(t)+{\tau}_{ij}(t) 
\label{model}
\end{equation}
where $\mu_{i}(t)$ is the mean process that describes the effects of
drug $i$ 
and ${\tau}_{ij}(t)$ is a zero-mean GP that induces temporal
correlation within a patient. 

Denoting the latent variable $a_{ijk}\equiv a_{ij}(t_k)$, we assume $e_{ijk}=I(a_{ijk}>a_{h})$, i.e.,
$$
e_{ijk} = \left\{ \begin{array}{rl}
 1 &\mbox{ if $a_{ijk}>a_h$} \\
  0 &\mbox{ otherwise}
       \end{array} \right.,
$$
where $a_h$ is an arbitrary threshold and $I(\cdot)$ is the
indicator function.  

In \eqref{model}, the choice of $\mu_i(t)$ depends on the underlying disease and drug mechanism. Based on prior knowledge, for
example, if we believe that the efficacy of a drug is
increasing over time, we could simply use a linear function $\mu_i(t)=\beta_{i0}+\beta_{i1}t$ with a positive slope; if we
believe the efficacy increases first then decreases due to drug
resistance, we could use a quadratic function. To accommodate
different shapes of response curves, we choose the polynomial regression model to describe
the drug mean effects over time, i.e.,
\begin{eqnarray}
\mu_{i}(t)=\beta_{i0}+\beta_{i1}t+\dots+\beta_{i,m_i}t^{m_i},
\label{eq:mean}
\end{eqnarray}
where
${\bm \beta}_i=(\beta_{i0}, \beta_{i1}, \dots, \beta_{i,m_i})'$ are regression
coefficients of time for each of the two treatment arms ($i=1, 2$). In
practice, the degree $m_i$ of the polynomial is unknown, and we let
$m_i=\{0, 1, \dots, M\}$ be a random variable taking integer values
between 0 and a large number $M$. It is important to allow
$(m_i, {\bm \beta}_i)$ to vary for different arms $i$ so that the drug
effects can be easily compared by posterior inference using these
parameters. Also, we do not consider nonparametric functions for
modeling $\mu_i(t)$ to avoid potential identifiability issue in the
model since $\tau_{ij}(t)$ is already nonparametric. 

Note that ${\bm \beta}_i$ describes the global response pattern of
each arm $i$. For the dependence across time within each
patient $j$, we assume that $\tau_{ij}(t)$ follows a zero-mean GP that induces the cyclic correlation of $a_{ij}(t)$. Let
\begin{equation}
 \tau_{ij}(t)\mid\Theta\sim GP\left(\bO_{K_{ij}}, \bC(\bt_{ij}; \Theta)\right),
\label{eq:cov}
\end{equation}
where $\bO_{K_{ij}}$ is a $K_{ij}$-dimension 0-vector, and $\bC(\bt_{ij}; \Theta)$ is a $K_{ij}\times K_{ij}$ covariance matrix indexed by $\Theta$.

Based on (\ref{model}), (\ref{eq:mean}) and (\ref{eq:cov}), the proposed latent Gaussian process functional regression model is given by
\begin{equation}
\small
P({\bm a}^K\mid {\bm \beta}, \bmm,  \Theta) \propto \prod_{i=1}^2\prod_{j=1}^{N_i}|\bC(\bt_{ij})|^{-\frac{1}{2}}\exp\left\{-\frac{1}{2}({\bm{a}}^K_{ij}-\bX_{ij}{\bm \beta}_i)'\bC(\bt_{ij})^{-1}({\bm{a}}_{ij}^K-\bX_{ij}{\bm \beta}_i)\right\},
\label{latent}
\end{equation}
where ${\bm{a}}^K_{ij}=(a_{ij1}, \dots, a_{ijK_{ij}})'$, $\bbeta=\{\bbeta_i\}$, $\bmm=\{m_i\}$, $i=1, 2$ and the design matrix $\bX_{ij}$ of the polynomial (\ref{eq:mean}) is

\[
\bX_{ij}=
 \begin{pmatrix}
  1 & t_1&\cdots &t_1^{m_i} \\
  1 & t_2&\cdots &t_2^{m_i}  \\
   \vdots  & \vdots &\ddots &\vdots  \\
  1 & t_{K_{ij}} &\cdots &t_{K_{ij}}^{m_i}  \\
 \end{pmatrix}.
\]
\
The covariance matrix $\bC(\bt_{ij})$ is usually parameterized to induce
within-patient dependence over time. The most popular parametrization \citep{Williams1998} of the stationary covariance
function $\bC(\bt_{ij}; \Theta)$ is a $K_{ij}\times K_{ij}$ matrix with the
$uv$-th element  $C_{uv}$ defined by
\begin{equation}
C_{uv}(\Theta)
=\theta_1^2\exp\left\{-r^2(t_u-t_v)^2\right\}+\delta_{uv}J^2, \quad
u,v = 1,\ldots, K_{ij}.
\label{coveq}
\end{equation}
In \eqref{coveq} $\Theta=(\theta_1, r)'$, $C_{uv}(\Theta)$ is the covariance of the GP for time
points $t_u$ and $t_v$, $\delta_{uv} = I(u=v)$,  and $J$ is the
variance on the diagonal reflecting the
amount of jitter \citep{Neal1998}, which usually takes on a small
value (e.g, $J=0.1$). This construction of the covariance matrix yields a strong correlation for observations at
time points close to each other 
and has been widely adopted \citep{Williams1998}.

\begin{figure}[ht!]
\includegraphics[scale=1]{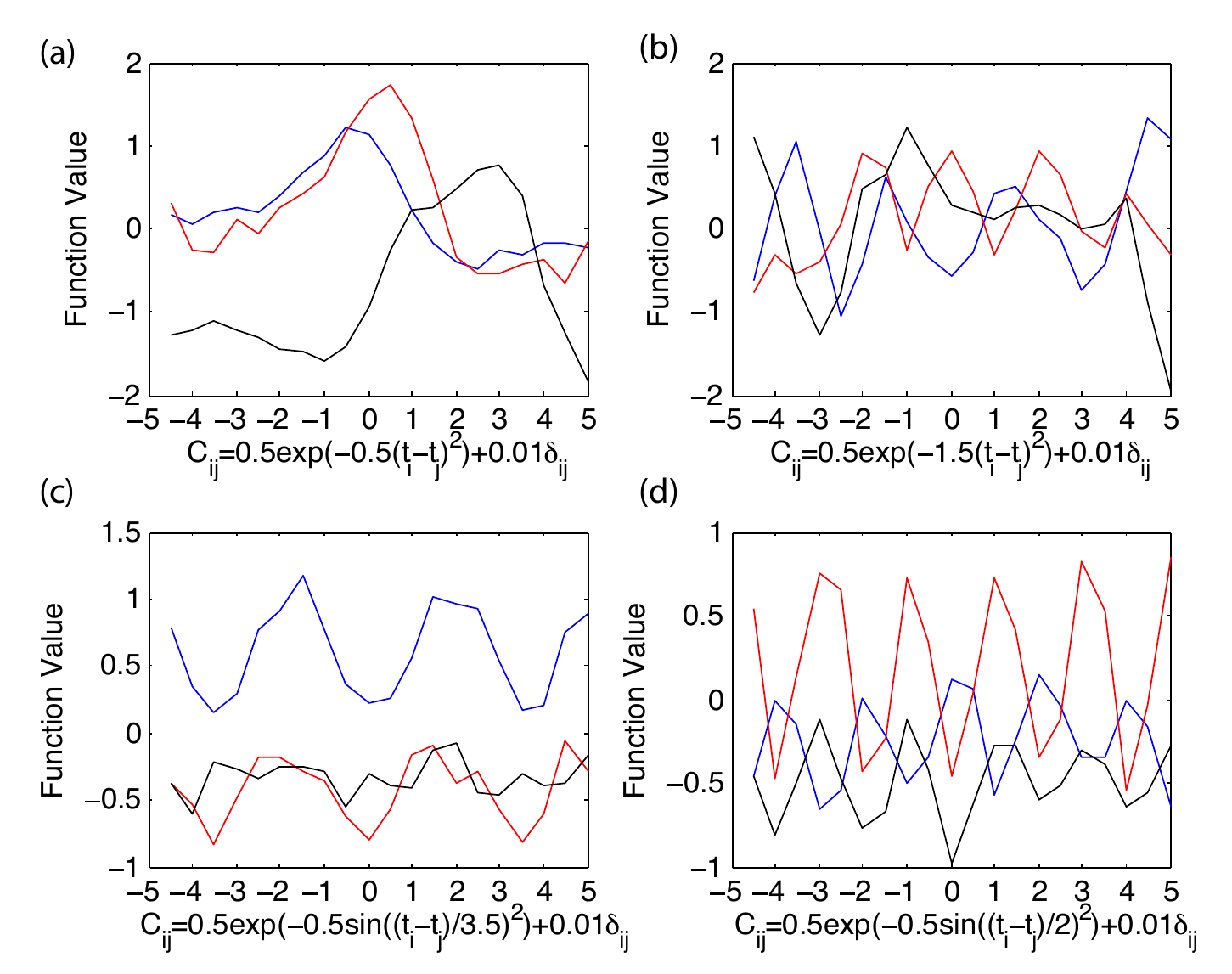}
\caption{Three random samples drawn from the Gaussian process
  $\tau_{ij}$ with four different covariance matrices.  The corresponding covariance is given below each plot. The decrease in length scale from (a) to (b) produces more rapidly fluctuating functions. The periodic properties of the covariance function in (c) and (d) can clearly be seen. }
\label{cov}
\end{figure}

However, if the system is recurrent, which is the case here for the lupus trial,
formulation (\ref{coveq}) is not suitable since it is aperiodic. Instead, we should consider other forms of covariance functions with periodicity. For example, a periodic model with known wavelength $\theta_2$ is given by
\begin{equation}
C_{uv}(\Theta)
=\theta_1^2\exp\left\{-r^2\sin^2(\frac{\pi(t_u-t_v)}{\theta_2})\right\}+\delta_{uv}J^2,
\label{eq:period}
\end{equation}
and  $\Theta=(\theta_1, \theta_2, r)'$. This covariance function
provides  strong correlations not only for adjacent time points but
also for those separated from each other by a distance
$\theta_2$ or its multiplications. We will use (\ref{eq:period}) as the model for our trial applications. As an illustration, 
Figure \ref{cov} shows some realizations of $\tau_{ij}(t)$ with different
covariance functions $\bC$. The realizations display some
periodic patterns although their shapes can be quite different. This
implies that the class of GP  models  with covariance \eqref{eq:period}
is general and can describe curves with a variety of different
shapes. 

Define the column vector $\ba^K=\{a_{ijk},  i=1,2; ~j=1,2, \dots,
 N_i; ~ k=1, 2,
\dots, K_{ij}\}^\prime$. The likelihood function is given by 
\begin{eqnarray}
P(\eb^K\mid \ba^K)&=&\prod_{i=1}^{2} \prod_{j=1}^{N_i}\prod_{k=1}^{K_{ij}}P(e_{ijk}\mid a_{ijk}) \nonumber\\
&=&\prod_{i=1}^{2} \prod_{j=1}^{N_i}\prod_{k=1}^{K_{ij}}\left\{I(a_{ijk}>a_h)I(e_{ijk}=1)+I(a_{ijk}<=a_h)I(e_{ijk}=0)\right\}.\nonumber\\
\label{sampling}
\end{eqnarray}

As a summary, the hierarchical model factors as
\begin{equation}
 P({\bm{e}}^K,  {\bm{a}}^K, \bmm, {\bm \beta},  \Theta) 
\propto \underbrace{P({\bm{e}}^K\mid{\bm{a}}^K)}_{\text{(\ref{sampling})}}\underbrace{P({\bm{a}}^K\mid {\bm \beta},  \bmm,  \Theta)}_{\text{(\ref{latent})}}P({\bm \beta}\mid \bmm)P(\bmm)P(\Theta),
\label{eq:joint}
\end{equation}
where $\bmm=(m_1, m_2)'$ and $\bm{\beta} =(\bm{\beta_1}',
  \bm{\beta_2}')'$.  The first factor is the sampling model (\ref{sampling}). The second factor is the
latent Gaussian process functional regression model
(\ref{latent}). The remaining factors are priors of $\bmm, {\bm
  \beta}$ and $(\theta_1, \theta_2, r)$. 
We assume a uniform prior on $P(m_i)$. For example, given a large integer $M$, $P(m_i)=1/(M+1)$ for $m_i=0, 1, \dots, M$. We assume
$${\bm \beta}_i \sim \text{Gaussian}({\bm \mu}_0^{m_i+1}, \sigma_0^2{\bf I}_{m_i+1})  $$
$$ \theta_1 \sim \text{Gaussian} (\mu_1, \sigma^2_1), \; r \sim
\text{Gaussian} (\mu_2, \sigma^2_2), \; \theta_2 \sim \text{Gaussian} (\mu_3, \sigma^2_3),$$
where ${\bm \mu}_0^{m_i+1}$ is $(m_i+1)$-dimension mean vector, and
${\bf I}_{m_i+1}$ is an $(m_i+1)\times (m_i+1)$ dimension identity
matrix.  We assume vague priors for ${\bm \beta}_i, \theta_1, r,
\theta_2$ by imposing large values of variances $(\sigma_0^2,
\sigma_1^2, \sigma_2^2, \sigma_3^2)$.

We will show that the LGP model in \eqref{eq:joint} possesses desirable
theoretical and numerical properties, such as consistency, the ability
to forecast for individual patients and to stop trials early. 

\subsection{MCMC Simulations}
The most significant computational cost of implementing a GP model
when dealing with a vast number of time points 
is the inversion of the covariance matrix $\bC$. Such an inversion is
required to make any predictions and, most significantly, in Bayesian
inference to evaluate the gradient of the log posterior distribution
over $\Theta$ in every Markov chain Monte Carlo (MCMC) iteration.

We carry out posterior inference based
on MCMC simulation. The proposed Gibbs sampler proceeds by iterating over the following transition probabilities
$$ P(\ba^K  \mid{\bm{e}}^K, {\bm \beta}, \bmm, \Theta), P(\bmm \mid {\bm a}^K,\Theta),  P( {\bm \beta}\mid {\bm{a}}^K, \bmm, \Theta) ,   P(\Theta\mid {\bm{a}}^K, {\bm \beta}, \bmm).$$
We start by generating ${\bm{a}}^K$ from their full conditional
posterior distributions -- truncated Gaussian. Efficient ways to sample from truncated Gaussian can be found in  \cite{geweke1991efficient, liechty2010multivariate, pakman2013exact}. Through a cycle of
Gibbs steps we also generate random draws of $\bmm$, ${\bm \beta}$ and $\Theta$.
When updating $\bmm$, we sample from the full conditional posterior
distributions, marginalized with respect to ${\bm \beta}$ (See
 Appendix A). Given $\bmm$, sampling of ${\bm \beta}$ is straightforward because of the fixed dimensionality and Gaussianity. This avoids the need for trans-dimensional MCMC. 
 The challenging step is to sample $\Theta$, the full conditional of
 which is analytically intractable. However, we can obtain the
 approximations using the most probable values such as Evidence
 maximization \citep{mackay1992practical} or other non-Gibbs MCMC samplers.  The most commonly used and
 straightforward method is the random-walk Metropolis
 sampling, but  \cite{Neal1998} argued that simple Metropolis algorithms
 were not efficient  
 and therefore advocated for the use of hybrid Monte Carlo \citep{duane1987hybrid}. This idea was implemented by  \cite{Barber97gaussianprocesses} who analyzed 
 classification problems and achieved better results using hybrid
 Monte Carlo compared to Metropolis algorithms. With the same
 motivation, we propose a hybrid MCMC algorithm to sample the
 hyperparameters $\Theta$ in the covariance function
 $\bC(\Theta)$. The idea behind hybrid MCMC is to augment the
 parameter space and draw Metropolis proposals with improved
 samplers. Suppose we want to draw Monte Carlo samples from a
 proposed density $P(\Theta)\propto \mathrm{exp}\{-E(\Theta)\}$, where
 $\Theta=(\theta_1, \dots, \theta_n)$. In physics terminology,
 $\Theta$ can be regarded as a position vector and $E(\Theta)$ the
 potential energy function. As a data augmentation step, 
 we then introduce another set of variables called ``momentum variables'', $\bw=\{w_1, w_2, \dots, w_n\}$, one $w_i$ for each $\theta_i$, with a kinetic energy function, $P(\bw)=\frac{1}{Z_K}\mathrm{exp}(-K(\bw))=(2\pi)^{-n/2}\mathrm{exp}(-\frac{1}{2}\sum_iw_i^2)$. The momentum variables are introduced to make random walks continue in a consistent direction until a region of low probability is encountered. The total energy function, known as Hamiltonian, is $H(\Theta,\bw)=E(\Theta)+K(\bw)=E(\Theta)+\frac{1}{2}\sum_iw_i^2$. The canonical distribution defined by this energy function is $P(\Theta,\bw)=\frac{1}{Z_H}\mathrm{exp}\{-H(\Theta,\bw)\}=P(\Theta)P(\bw)$. 
Random samples of $(\Theta, \bw)$ are jointly proposed by appropriate
probabilities (see Appendix A for details). Samples from the marginal
distribution for $\Theta$  can be obtained by ignoring the values
$\bw$. Specifically, we use a discretized approximation called the
{\it leapfrog} algorithm \citep{Neal1997}, details of which are again described in Appendix A.

To summarize, the proposed MCMC algorithm includes a Gibbs sampler for
${\bm{a}}^K$, $\bmm$,  ${\bm \beta}$ and a hybrid Monte Carlo algorithm for sampling
$\Theta$.   The algorithm converges well with an initial burn-in of 2,000
iterations and a total of 10,000 iterations with a lag of 10 in our simulation
studies. We conduct convergence diagnostics using R package {\it coda} and find no evidence for convergence problems.  Traceplots and empirical autocorrelation plots (not shown) for the imputed parameters indicate a well mixing Markov chain. 

\section{Posterior Consistency}
\label{sec:post}
We discuss theoretical properties of the LGP model and show
that it is consistent as the number of treated patients 
goes to infinity. For simplicity in the following discussion, we only consider one treatment arm by dropping index $i$ and use $n$ to denote the number of patients and $K_j$ to denote the number of time points observed for patient $j$. The proposed LGP model can be summarized as follows: 
\begin{eqnarray}
e_{jk}=e_j(t_k)\mid p(t_k) &\sim& \mathrm{Bernoulli}(p(t_k)), \ \ k=1, \dots, K_j, \ j=1, \dots, n\nonumber\\
p(t) &=&H\left(a(t)\right)\equiv Pr\left(a(t)>a_h \right),  t \in
[0, T_E]  \nonumber\\
a(\cdot) &\sim& GP(\mu(\cdot), C(\cdot, \cdot)), 
\label{modeleq}
\end{eqnarray}
where observation times $t_k$'s are irregularly spaced points for each patient and fixed in advance on the compact set $\cT=[0,
T_E]\subset \mathcal{R}$ in which $T_E$ is the length of the trial, and
$H$ is a known, strictly increasing, Lipschitz continuous cumulative
function. Note that $a(\cdot)$ is a Gaussian process
parameterized by its mean function $\mu$: $\cT\rightarrow \mathcal{R}$
and its covariance function $C: \cT^2\rightarrow\mathcal{R}$, denoted by
$GP(\mu, C)$. Additionally, we assume that the underlying true
probability  function is $p_0(t)=H(a_0(t))$, and $a_0(t)$, the true GP, 
has a continuously differentiable sample path on $\cT$. 
 Without loss of generality,
 we assume a linear mean function $\mu(t)=\beta_0+\beta_1t$. The proof for the case of $m$-polynomial
is easily extended.  In addition, we assume that the true covariance
function has the form $C_0(t_u, t_v; \Theta)=C_0(|t_u-t_v|)$, where
$C_0$ is a positive multiple of a nowhere-zero density function of one
real variable. In our model, we consider a simple form of covariance
function (\ref{eq:period}) without the jitter part. Then
$C_0(t)=\theta_1^2\mathrm{exp}\left\{-r^2\mathrm{sin}^2(\frac{\pi
}{\theta_2}t)\right\}$. We rewrite $C_0(t) \equiv C_0(t;~\theta_1, r,
\theta_2)$ to prepare for the coming discussion.

Posterior consistency using GP priors has been
extensively studied, see \cite{choudhuri2004bayesian},
\cite{choi2004posterior,choi2007posterior},
\citet{ghosal2006posterior}, and  \citet{tokdar2007posterior}, among
others. However,
little is known about the type of LGP models proposed in this paper.  We first present posterior consistency results 
based on a general modeling framework under a wide class of GP priors $a(\cdot)$ and link functions $H$, which can be any strictly increasing, Lipschitz continuous function.
Our proposed LGP model in \eqref{modeleq} is simply a special case
of the general results. 
The proofs of all lemmas and theorems stated in this section are given in Appendix B.

 To start,  let $\mathcal{F}$ be the set of Borel
measurable functions defined on $\mathcal{R}$, the real line. For now, assume that we
have a topology on $\mathcal{F}$ that measures the distance
between any two functions in $\mathcal{F}$. We also follow \cite{ghosal1999posterior}, to
define a probability measure on the functional space. That is, we
operate under the probability space $(\mathcal{F},
\sigma_{\mathcal{F}}, P_{\mathcal{F}})$, where $\sigma_{\mathcal{F}}$
is the $\sigma$-field of $\mathcal{F}$ and $P_{\mathcal{F}}$ is the
probability measure for functions. 
Suppose $p(t)\in\mathcal{F}$, and let $P_p$ stand for the
probability measure corresponding to $p(t)$ on the probability space
for the real line, $(\mathcal{R}, \mathcal{B}, P)$, where $\mathcal{B}$ and $P$
are the Borel set and the probability measure
on the real line, respectively. To clarify, we use $(\mathcal{F},
\sigma_{\mathcal{F}}, P_{\mathcal{F}})$ and $(\mathcal{R},
\mathcal{B}, P)$ to denote the probability spaces for random functions
and random variables, respectively. Denote $K=\sum_{j=1}^nK_j$. 
Dropping index $j$ we 
let $(e_{t_1}, \dots,
e_{t_K})$ be the collection of $K$ observed binary outcomes across $n$
patients, sorted over time. Under \eqref{modeleq} each $e_{t_k} \mid
p(t_k) \sim Bern(p(t_k))$. 
We assign prior distribution $\Pi$ for $p(t)$ and the posterior
distribution given $e_{t_1}, \dots, e_{t_K}$, denoted by
$\Pi(\cdot\mid e_{t_1}, \dots, e_{t_K})$. We will show consistency of the posterior distribution of $p$  under 
some regularity conditions, 
 if
$e_{t_k}$'s are generated conditional on the true function $p_0(t)$,
i.e., $e_{t_k} \mid p_0(t_k) \sim Bern(p_0(t_k))$.

Below we introduce the metric on $\mathcal{F}$, three lemmas, and the
main theorem. 

\bigskip

\begin{definition}
Let $\mathcal{F}$ be the set of Borel
measurable functions defined on $\mathcal{R}$. For any $f_0\in \mathcal{F}$, denote the
Kullback-Leibler neighborhood $KL_{\epsilon}(f_0)=\{f: \int
f_0log(f_0/f)<\epsilon\}$. Let $\Pi$ be a prior on $\mathcal{F}$, we say $f_0$ is in the K-L support of $\Pi$ if $\Pi(KL_{\epsilon}(f_0))>0$ for all $\epsilon>0$. 
\end{definition} 

\bigskip

\begin{lemma}
\label{distance}
Let $([0, T_E], \mathfrak{B}, P)$ be a probability space on the
real line, and let
$\mathcal{F}$ be the set of all real-valued Borel measurable functions $f$: $[0, T_E] \rightarrow [0,1]$. Let $P(t)=l(t)/T_E$ be a probability measure, where $l(t)$ is the Lebesgue measure on $[0, T_E]$. Define
$$d(f, g)=\mathrm{inf}\{\epsilon: P(\{t: |f(t)-g(t)|>\epsilon\})<\epsilon\}.$$
Then $d(\cdot, \cdot)$ induces a metric, and $f_K$ converges to $f$ in probability if and only if $\mathrm{lim}_{K\rightarrow\infty}d(f_K, f)=0$.
\end{lemma}

\vspace{.1in}
\begin{lemma}
\label{th:inequality}
Let $0<\epsilon_0<\frac{1}{2}$ and $\epsilon_0<a, b<1-\epsilon_0$. Then there exists a constant $L$ depending on $\epsilon_0$ such that
$$a\log\frac{a}{b}+(1-a)\log\frac{1-a}{1-b}\leq L(a-b)^2. $$
\end{lemma}

\bigskip

\begin{lemma}
\label{gpineq}
Assume the compact supports of $\theta_1, r, \theta_2$ are $B_1, B_2$
and $B_3$ respectively. Let $B=B_1\times B_2\times B_3$, a compact
support of $(\theta_1, r, \theta_2)$. Define two functions
$\rho_1(\theta_1, r, \theta_2)=C_0(0; \theta_1, r, \theta_2)$ and
$\rho_2(\theta_1, r, \theta_2)=-C''_0(0; \theta_1, r, \theta_2)$,
where $C''_0$ is the second derivative of $C_0$. 
Let $a(\cdot)$ be the GP on $\cT$ in (\ref{modeleq}), where $\cT$ is a bounded subset of $\mathcal{R}$. Then $a(\cdot)$ has differentiable sample paths and the derivative process $a'(\cdot)$ is also a GP. Further, there exist constants $A$ and $d_1, d_2$ such that
\begin{eqnarray*}
\mathrm{sup}_{(\theta_1, r, \theta_2)\in B}Pr\left\{\mathrm{sup}_{t\in [0, T_E]}|a(t)|>M_n\mid \theta_1, r, \theta_2\right\}\leq Ae^{-d_1n} \nonumber\\
\mathrm{sup}_{(\theta_1, r, \theta_2)\in B}Pr\left\{\mathrm{sup}_{t\in [0, T_E]}|a'(t)|>M_n\mid \theta_1, r, \theta_2\right\}\leq Ae^{-d_2n}, \nonumber
\end{eqnarray*}
where $M_n=O(n^{1/2})$.
\end{lemma}

\bigskip

Next we present the main results in Theorem \ref{consistency}. 
\cite{choi2007posterior} proposed a Consistency Theorem (see the Appendix C) as an
extension of Schwartz's theorem \citep{schwartz1965bayes} to
independent but non-identically distributed cases. We make use of such
an extension to achieve the posterior consistency for our proposed
LGP by verifying the two conditions in \cite{choi2007posterior}.

\bigskip

\begin{theorem}
\label{consistency}
Suppose that $K/n = O(1)$, i.e., when the number of patients $n$ goes to
infinity, the total number of distinct observational time points
$K=\sum_{j=1}^n K_j $ in $[0, T_E]$ also goes to
infinity. Assume
that the mean function $\mu(\cdot)$ and $C(\cdot, \cdot)$ in
the Gaussian process prior (\ref{modeleq}) satisfy the following conditions:
\begin{enumerate}
\item $C(\cdot, \cdot)$ is nonsingular,
\item $C(t_u, t_v)$ has the form $C_0(|t_u-t_v|)$, where $C_0(t)$ is a positive multiple of a nowhere zero density function on $\mathcal{R}$ and four times continuously differentiable on $\mathcal{R}$,
\item the mean function $\mu(t)$ is continuously differentiable in
  $[0, T_E]$, and 
\item there exist $0<\delta<\frac{1}{2}$ and $b_1, b_2>0$ such that
$$Pr(\theta_1^2>n^{\delta})<b_1\mathrm{exp}(-b_2n), Pr(r^2>n^{\delta})<b_1\mathrm{exp}(-b_2n),$$
$$Pr(\theta_2^2<n^{-\delta})<b_1\mathrm{exp}(-b_2n).$$
\end{enumerate}

\noindent Let $P_0^K$ denote the joint distribution of $\{e_{t_k}\}_{k=1}^K$
assuming that $p_0(t)=H(a_0(t))$ is the true
 probability function, where 
 $a_0(t)$ is the true GP that  has a continuously differentiable sample path on $\cT$.    Then for every $\epsilon>0$, letting $S^1_\epsilon(p_0) = \{p \in
\mathcal{F}: d(p, p_0) > \epsilon\}$, as $n \to \infty$ 
\begin{eqnarray}
\label{consistency1}
\Pi\{S^1_\epsilon(p_0) \mid e_{t_1}, \dots, e_{t_K}\} \rightarrow 0 \ \   \ [P_0^{\infty}]  
\end{eqnarray}
In addition, for every $\epsilon>0$, letting $S^2_\epsilon(p_0) = \{p \in
\mathcal{F}: \int |p(t)-p_0(t)|dt>\epsilon \}$, as $n \to \infty$ 
\begin{eqnarray}
\label{consistency2}
\Pi\{ S^2_\epsilon \mid e_{t_1}, \dots, e_{t_K}\}\rightarrow 0 \ \   \ [P_0^{\infty}]
\end{eqnarray}
\end{theorem}

In general, if more hyperparameters are introduced in the
covariance function, we can verify posterior consistency by assuming
compact supports of hyperparmaters and proper continuity conditions as
in Theorem \ref{consistency}.
  
\section{Posterior Inference}
\label{sec:stop}
\subsection{Forecast}
An important and useful feature of the LGP model is the
ability to forecast,
i.e., making predictions on $e_{ij}(t_{K_{ij}+s})$ for future time points
$t_{K_{ij}+s}$, $s=1, 2, \dots, S_{ij}.$ The forecast uses the
posterior predictive distribution
$p\{e_{ij}(t_{K_{ij}+s})\mid \eb^K\}$, where $\eb^K$ are observed
data at $K_{ij}$ past time points. In particular, define the
posterior predictive probability of response as 

\begin{eqnarray}
\label{forecast}
q_{ijs} &\equiv & ~\hskip-0in Pr\left\{e_{ij}(t_{K_{ij}+s})=1 \mid
  {\bm{e}}^K\right\} = Pr\left\{a_{ij}(t_{K_{ij}+s}) > a_h \mid
  {\bm{e}}^K \right\} \nonumber\\ 
  &=&    \int I\left\{a_{ij}(t_{K_{ij}+s})>a_h\right\}~p\left\{a_{ij}(t_{K_{ij}+s})\mid{\bm
   a}_{ij}^K, m_i, {\bm \beta}_i, \Theta\right\} \nonumber\\
   &&\times ~ p\left({\bm
   a}_{ij}^K, m_i, {\bm \beta}_i, \Theta\mid{\bm e}^K\right)  ~da_{ij}(t_{K_{ij}+s})~d{\bm a}_{ij}^K~dm_i~d{\bm \beta}_i
 ~d\Theta. 
\end{eqnarray} 

\noindent The key component in $q_{ijs}$ is the conditional distribution $p\left\{a_{ij}(t_{K_{ij}+s})\mid{\bm a}_{ij}^K, m_i, {\bm \beta}_i, \Theta\right\}$, for all future
points $t_{K_{ij}+s}$, $s=1, 2, \ldots, S_{ij}$. Therefore, we consider the
vector ${\bm a}^{(S-K)}_{ij}= \left(a_{ij}(t_{K_{ij}+1}),
a_{ij}(t_{K_{ij}+2}), \dots, a_{ij}(t_{K_{ij}+S_{ij}}) \right)^\prime$,  the latent
Gaussian variables for future time points. 
By definition of the GP model,  the joint distribution of
$p\left\{{\bm a}^K_{ij}, {\bm a}^{(S-K)}_{ij} \mid m_i, {\bm \beta}_i,
  \Theta \right\}$ is a multivariate Gaussian, and so is the
conditional $p\left\{a_{ij}(t_{K_{ij}+s})\mid{\bm a}_{ij}^K, m_i, {\bm
    \beta}_i, \Theta\right\}$. Denote its conditional CDF by $\Phi_{S-K|K}\left(\cdot \mid {\bm a}_{ij}^K, m_i, {\bm
    \beta}_i, \Theta \right)$. Then the
Monte Carlo estimate of $q_{ijs}$ is given by

\begin{equation}\label{qhat}
  \hat{q}_{ijs} = \sum_{b=1}^B \frac{1-\Phi_{S-K|K}\left(a_h\mid {\bm
      a}_{ij}^{K, (b)}, m_i^{(b)}, {\bm \beta}_i^{(b)}, \Theta^{(b)}\right)}{B},
\end{equation}
where superscript $^{(b)}$ denotes the MCMC sample of the $b$-th iteration
and $B$ is the total number of MCMC samples kept for posterior
inference. 

Note that the forecast is subject-$j$-specific. That is, based on the
observed outcomes for patient $j$, the forecast could be different due
to the subject-specific $\tau_{ij}$ in our original LGP model
\eqref{model}. This is useful in practice allowing individual patients
and their physicians to prepare for future disease relapses. 

\subsection{Trial Monitoring Rules}
\label{stoprule}
In Section \ref{probabilitymodel},  we assume that the treatment mean
effect for arm $i$ is $\mu_{i}(t)=\beta_{i0}+\beta_{i1}t+\dots+\beta_{i,m_i}t^{m_i}$ with a random degree $m_i$. 
For example, when $m_i=1$ and slope $\beta_{i1}$ is positive, the mean
effect $\mu_i(t)$ 
is a simple linear function
increasing with time; if $m_i=2$ with $\beta_{i2}<0$, the mean
effect $\mu_i(t)$ first increases and then decreases, potentially
due to drug resistance. Estimating $\bbeta_i$ for two arms 
allows for the comparison of 
the overall drug effects and hence trial monitoring.  
For example, if
the experimental arm is considered to perform no better than the
control arm in terms of increasing disease remission time, the trial
should be stopped early due to ethical and logistic reasons.

Denote the duration of the disease remission (DDR)  as 
\begin{equation}
  T_i (\bbeta_i, m_i) =l\{t: 
  \mu_i(t)>a_h\}, \label{eq:r-time}
\end{equation} 
where $l\{\}$ is the Lebesgue measure on the real line. That is, $T_i$ is
the length of time intervals in which $\mu_i(t)$ is above $a_h$ for
arm $i$, i.e., patients stay in remission. The arm with longer DDR is more desirable. 

The proposed monitoring criterion uses the posterior
probability 
\begin{eqnarray}
\eta &=& \mathrm{Pr}(T_2>T_1+\delta\mid\eb^K) \nonumber\\
&=&\int I\left\{T_2(\bbeta_2, m_2)>T_1(\bbeta_1, m_1)+\delta\right\}
~p(\bbeta, \bmm \mid \eb^K)~d\bbeta~d\bmm,
\end{eqnarray}
where $\delta \geq 0$ is a threshold that defines the desirable
increment of the efficacy outcome for the experimental arm over the
control. The value of $\delta$ is fixed by investigators, and should
reflect the minimum clinically meaningful improvement. A larger
$\eta$ value indicates a larger chance that treatment 2 is better at
keeping patients in remission than treatment 1.

The Monte Carlo estimate of $\eta$ is given by
\begin{equation}
\hat{ \eta}= \sum_{b=1}^B \frac{I\left\{{T}_2(\bbeta_2^{(b)}, m_2^{(b)})>{T}_1(\bbeta^{(b)}_1, m^{(b)}_1)+\delta\right\}}{B},
\end{equation}

 Let $\xi_U$ be an upper probability boundary above which the
 trial will be terminated early and the  experimental treatment
 declared superior if $\mathrm{Pr}(T_2>T_1+\delta\mid\eb^K)\geq
 \xi_U$. Similarly, let $\xi_L$ be a lower boundary below which the trial will be terminated early due to  futility if $\mathrm{Pr}(T_2>T_1+\delta\mid\eb^K)\leq \xi_L$.

Our proposed trial-monitoring rules are given as 
 $$
 \mathrm{Decision} = \left\{ \begin{array}{ll}
  \mathrm{stop \ and \  declare  \ arm \ 2 \  superior} &\mbox{ if
    $\hat{\eta} \geq \xi_U$, } \\
  \mathrm{continue \  enrolling \ patients} &\mbox{ if
    $\xi_L<\hat{\eta}<\xi_U$, } \\
 \mathrm{stop \ and \ declare \ arm \ 2 \  not \ superior} &\mbox{ if
   $ \hat{\eta} \leq \xi_L$.} \\
        \end{array} \right.
 $$
 Cutoffs $\xi_U$ and $\xi_L$ can be calibrated based on
 simulations. The values $\xi_U=0.95$ and $\xi_L=0.05$ provided
 desirable performance in the simulation for the lupus trial.

\section{Model Assessment via Sensitivity Analysis}
\label{simulation}
When making forecasts on $e_{ij}(t_{K_{ij}+s})$ for future time
points, three factors can affect the forecasting accuracy: 1) the
choice of the mean functions $\mu_i(t)$ and covariance functions
$\tau_{ij}(t)$, 
2) the number of observed time points $K_{ij}$ at the time of
forecast, and  3) the choice of
the threshold $a_h$. To evaluate the
impact of these model components, we perform a sensitivity
analysis based on simulation. To simplify the simulation setup, we only consider
data from one arm and focus on the accuracy of posterior
estimation and forecast instead
of trial characteristics. Therefore the index $i$ is dropped in
the
discussion within this section.

We assumed that the sample size was 100 patients for a single arm.  We first
generated values $\ba^K$ given $\mu(t)$ and $\tau_j(t)$ at a total of 35
time points $t_1, t_2, \dots, t_{35}$,  out of which the first
$K_j \leq 32$ time points were used to predict the responses of
patient $j$ at the
last three time points. 
 After generating $\ba^K$, we generated $\eb^K$ given the threshold
 $a_h$ of our choice. Without loss of generality, we assumed that all the patients had the same $K_j=K$ for $ j=1. \dots, 100$.   We specified various choices of $K=20, 23, 26,
 29, 32$ to study their effects on forecasting in each scenario. To
 show the robustness to the threshold $a_h$, we assumed $a_h=0$ when
 we generated the simulation data and set $a_h=0$ or $a_h=0.5$ when we
 fitted models to the simulated data.

We considered five scenarios. In
scenarios 1-3, we generated data by using different polynomial mean
functions $\mu(t)$ and the covariance function (\ref{eq:period}) with
$\theta_1=1, \theta_2=3.5$  and $r=2$.  For ease of exposition, we
display (\ref{eq:period}) again:  
$C_{uv}(\Theta)
=\theta_1^2\exp\left\{-r^2\sin^2(\frac{\pi(t_u-t_v)}{\theta_2})\right\}+\delta_{uv}J^2$. In
scenario 1, the true mean $\mu(t)=\beta_0$, where $\beta_0=-0.8$. In
scenario 2, the true $\mu(t)=\beta_0+\beta_1t$, where
$\beta_0=-0.8, \beta_1=0.4$. In scenario 3, the true 
$\mu(t)=\beta_{0}+\beta_{1}t+\beta_2t^2$, where $\beta_0=-1,
\beta_1=3.5, \beta_2=-1$. In words, scenarios 1-3 represent constant mean, linear mean, and quadratic mean respectively.
We fitted the LGP model (\ref{eq:joint}) 
to the simulated data and assumed vague priors ${\bm \beta}\sim \text{Gaussian}({\bm 0}, 10^2{\bf I})$,  $\theta_1, \theta_2, r \sim \text{Gaussian} (0, 10^2)$.

In scenario 4, we generated data from the trigonometric mean function
$\mu(t)=\alpha+\sin(\beta_0\pi t)$ with $\alpha=-0.8, \beta_0=1.5$, and
the covariance function  (\ref{coveq}) with $\theta_1=1$ and $r=3$. We
then fitted the model using the same trigonometric mean function but
with unknown
parameters, and the covariance
function (\ref{coveq}) to the simulated data and estimated unknown
parameters $\alpha, \beta_0, \theta_1$ and $r$, for which we also assumed
vague priors $\alpha, \beta_0, \theta_1, r \sim \text{Normal} (0,
10^2)$.

\begin{figure}[h!]
\begin{tabular}{cc}
 \includegraphics[width=.5\textwidth]{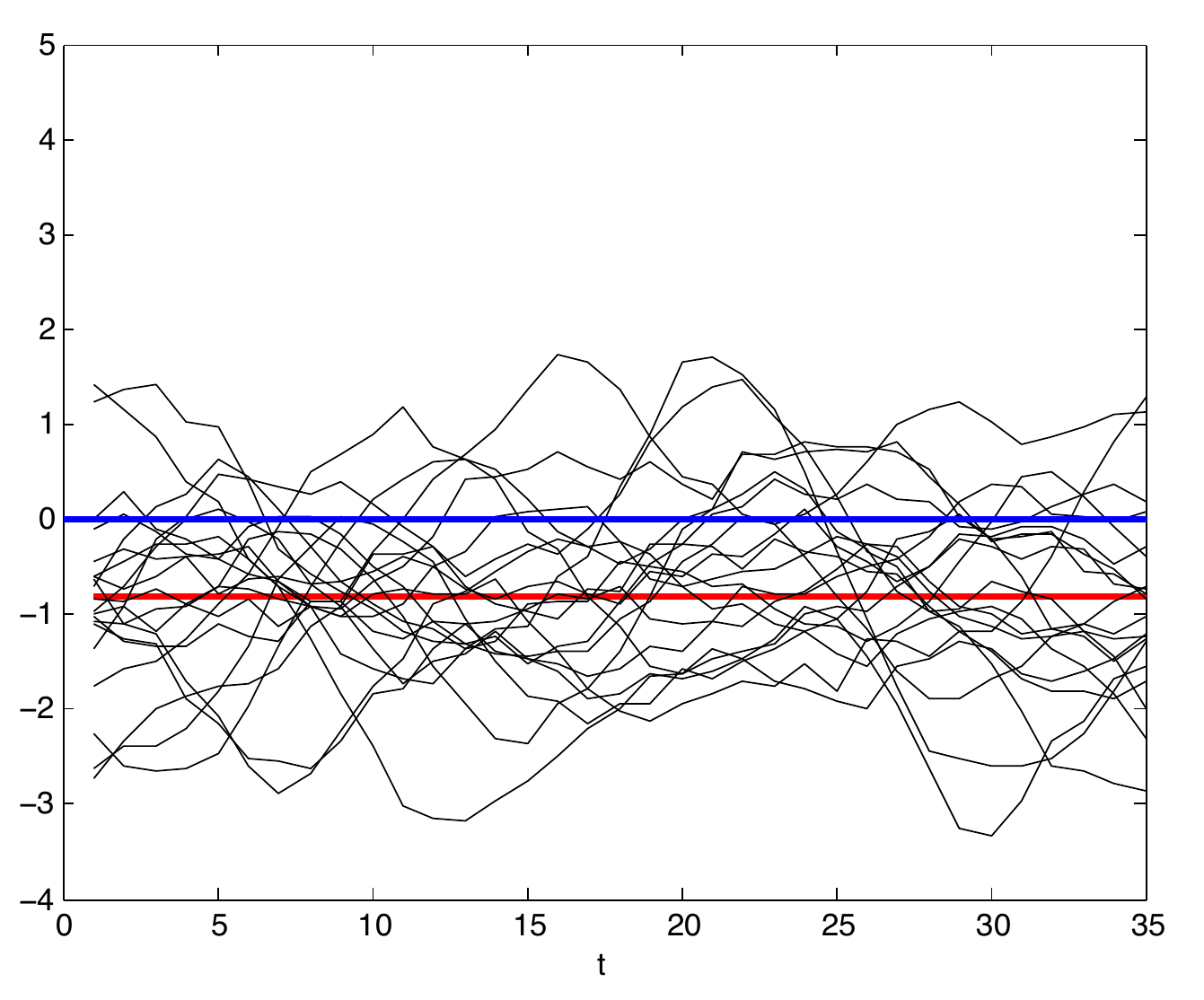}&
 \includegraphics[width=.5\textwidth]{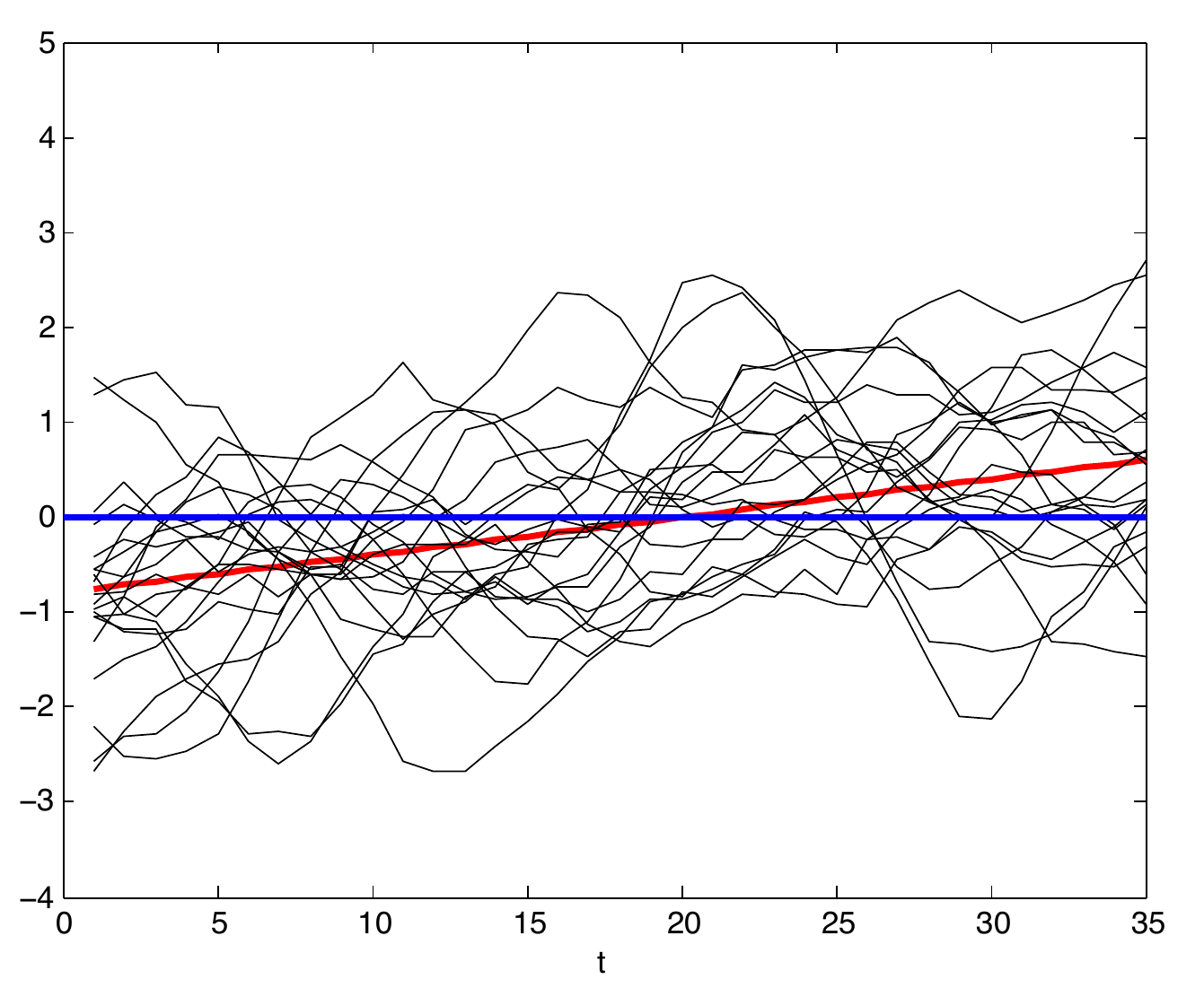}\\
 Scenario 1&
 Scenario 2\\
  \includegraphics[width=.5\textwidth]{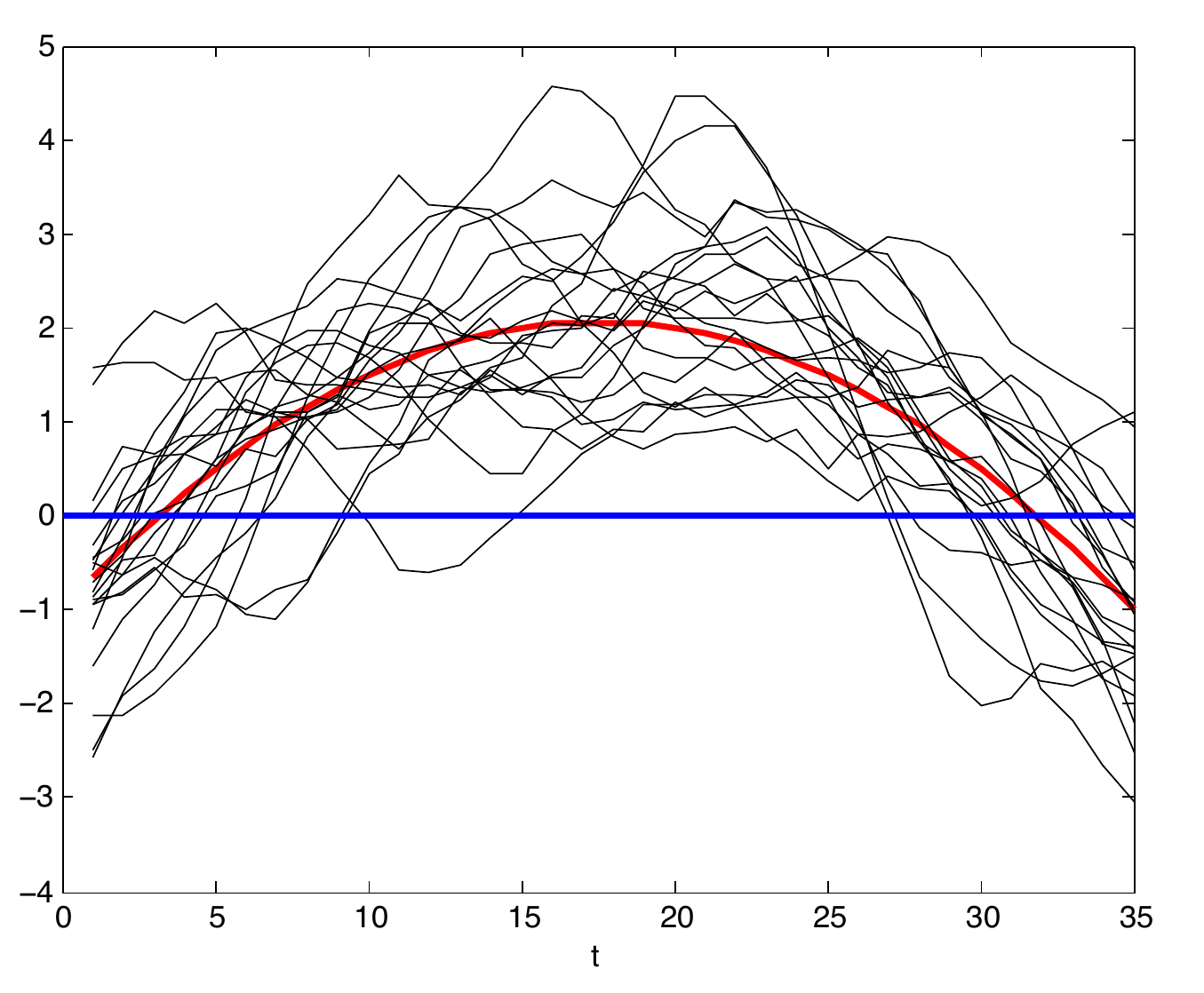}&
   \includegraphics[width=.5\textwidth]{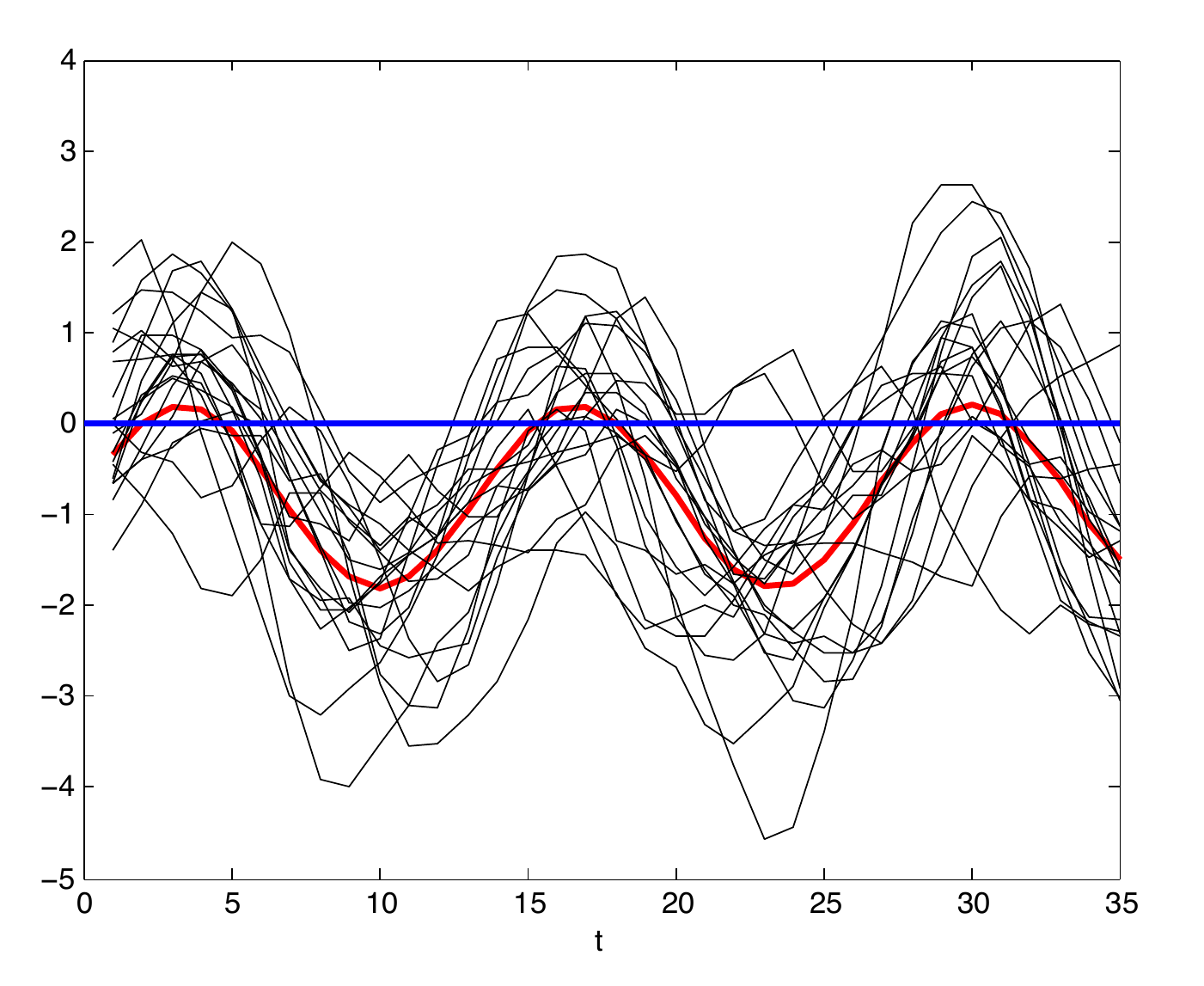}\\
 Scenario 3&
 Scenario 4\\
 \end{tabular}
 \caption{Sample curves of latent variable values for 20 randomly
 selected patients in scenarios 1-4. The red curve in each plot 
 represents the mean function $\mu(t)$ of the GP. The horizontal blue line represents the threshold $a_h$.}
 \label{fig:sample1}
 \end{figure}

In scenarios 1-4, we took periodicity into consideration by placing
the trigonometric function in the fitted model. Scenarios 1-3 had
a trigonometric function in the covariance function, while scenario 4
assumed a trigonometric mean function. To show the differences between
these two configurations, in Figure \ref{fig:sample1} we plotted
simulated $\ba$ for 20 randomly selected replications under scenarios
1-4. We can see that the overall mean in the top right panel increases with time and each patient curve has a different periodic behavior in scenario 2. In contrast, in the bottom right panel for scenario 4, the overall mean has an obvious periodic pattern and each individual curve has its own variability around the mean.

As discussed above, putting a trigonometric function in the mean
function $\mu(t)$ or covariance function $\tau_j(t)$ leads to different group and
individual behaviors. Since both models are periodic, 
we designed scenario 5 to investigate whether one model can predict for the
other. For this purpose, we generated data from the true model with
mean $\mu(t)=\sin(\beta_0\pi t)$, where $\beta_0=1$ and $\tau_j(t)$ from a GP with zero mean and covariance
function $C$ in (\ref{coveq}) with $\theta_1=1$ and
$r=3$, which does not have a trigonometric component. Then we fitted
model (\ref{eq:joint})  with a polynomial mean function and the covariance
function having trigonometric components. 

For each of the five scenarios described above, we specified five
 different values for $K=20, 23, 26$, $29, 32$,  and two threshold values
$a_h=0$ or $0.5$. Therefore, we obtained a total of
$5\times2\times5=50$ cases. For each case, we implemented
the proposed LGP model with 
10,000 MCMC iterations with a burn-in of 2,000 iterations, finished in 30 minutes.
The convergence of the MCMC algorithm was diagnosed by standard methods in
R package {\it{coda}}.  All the chains converged quickly and mixed
well. 

\begin{figure}[h!]
\begin{tabular}{cc}
 \includegraphics[width=.5\textwidth]{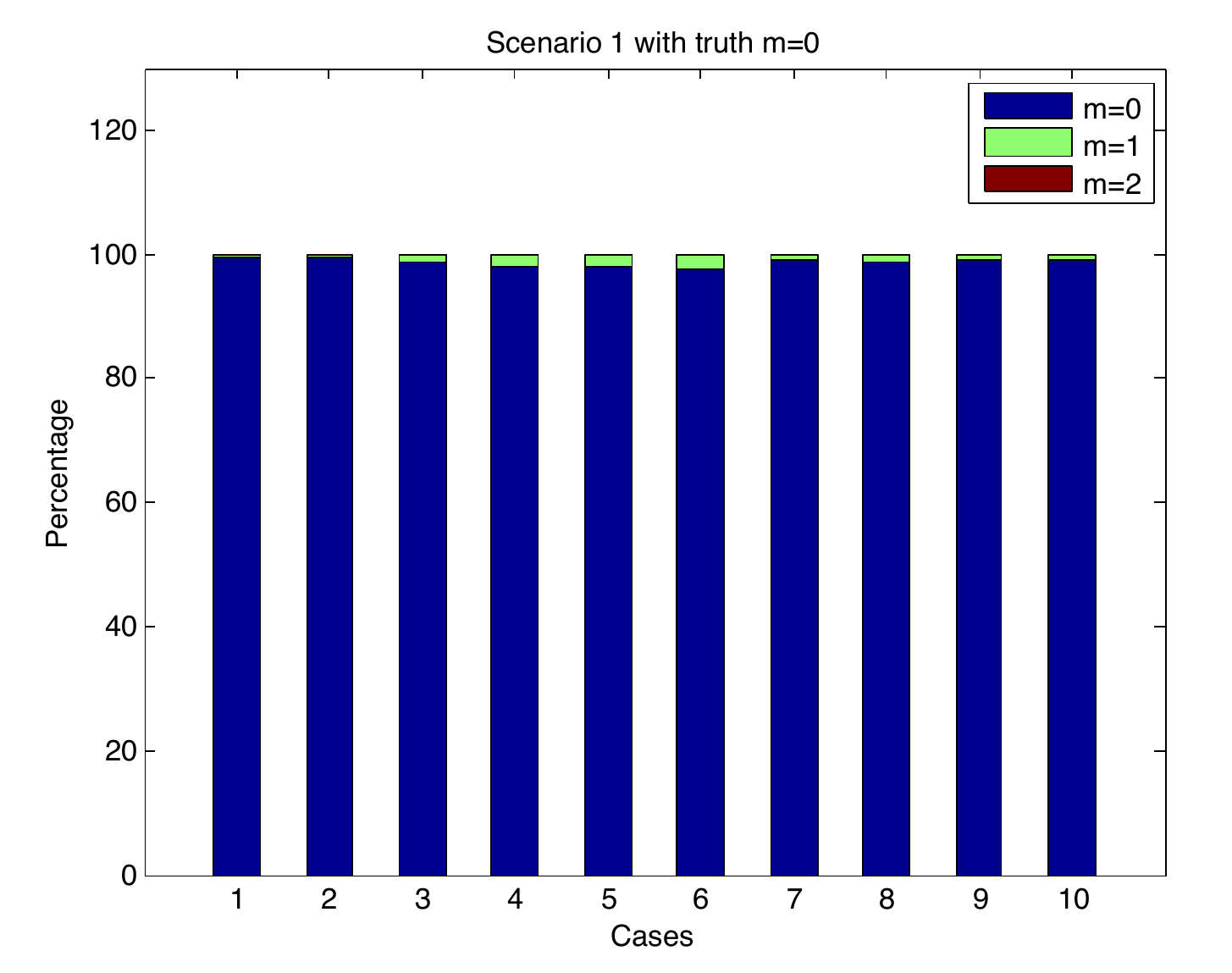}&
 \includegraphics[width=.5\textwidth]{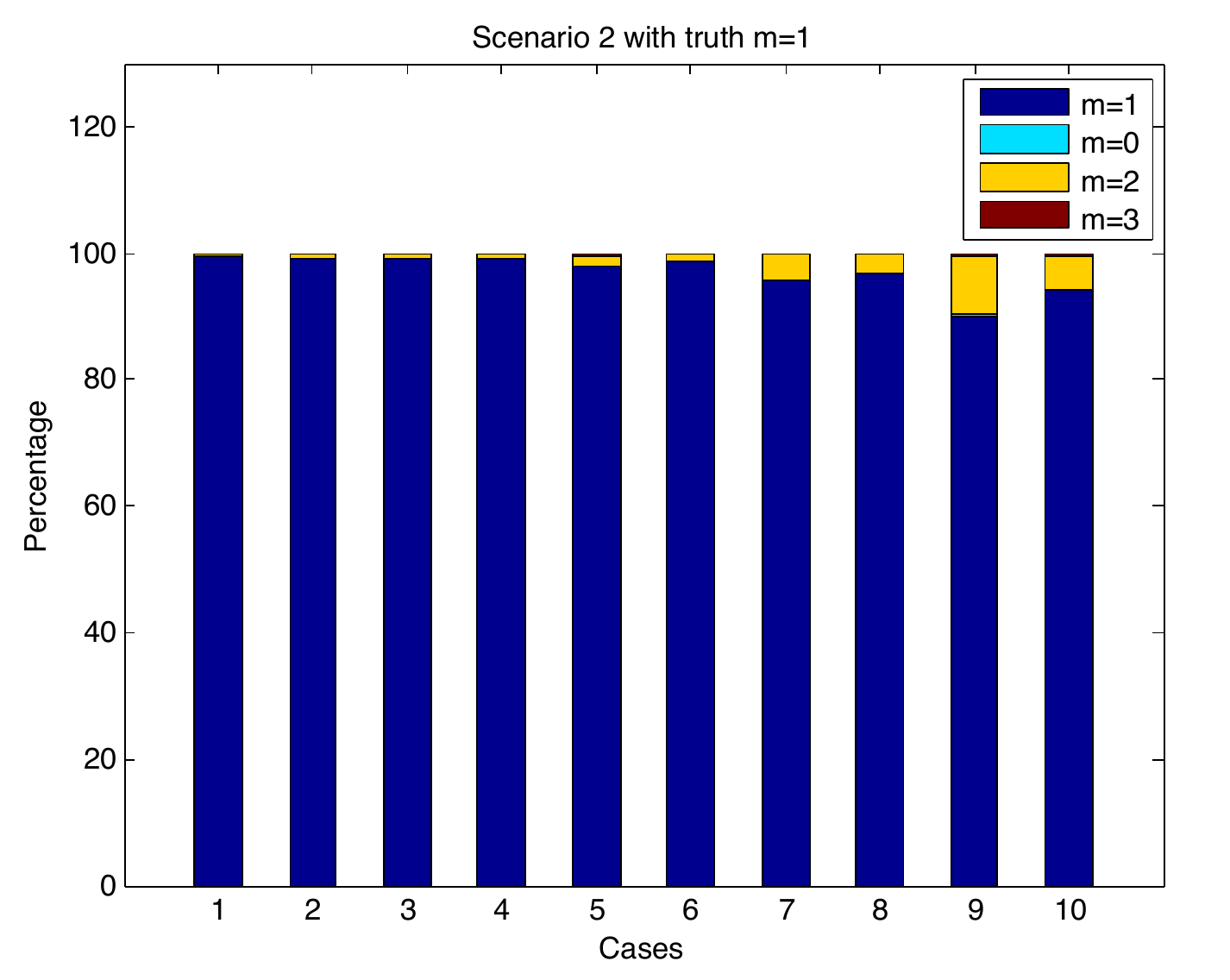}\\
  Scenario 1&
 Scenario 2\\
 \includegraphics[width=.5\textwidth]{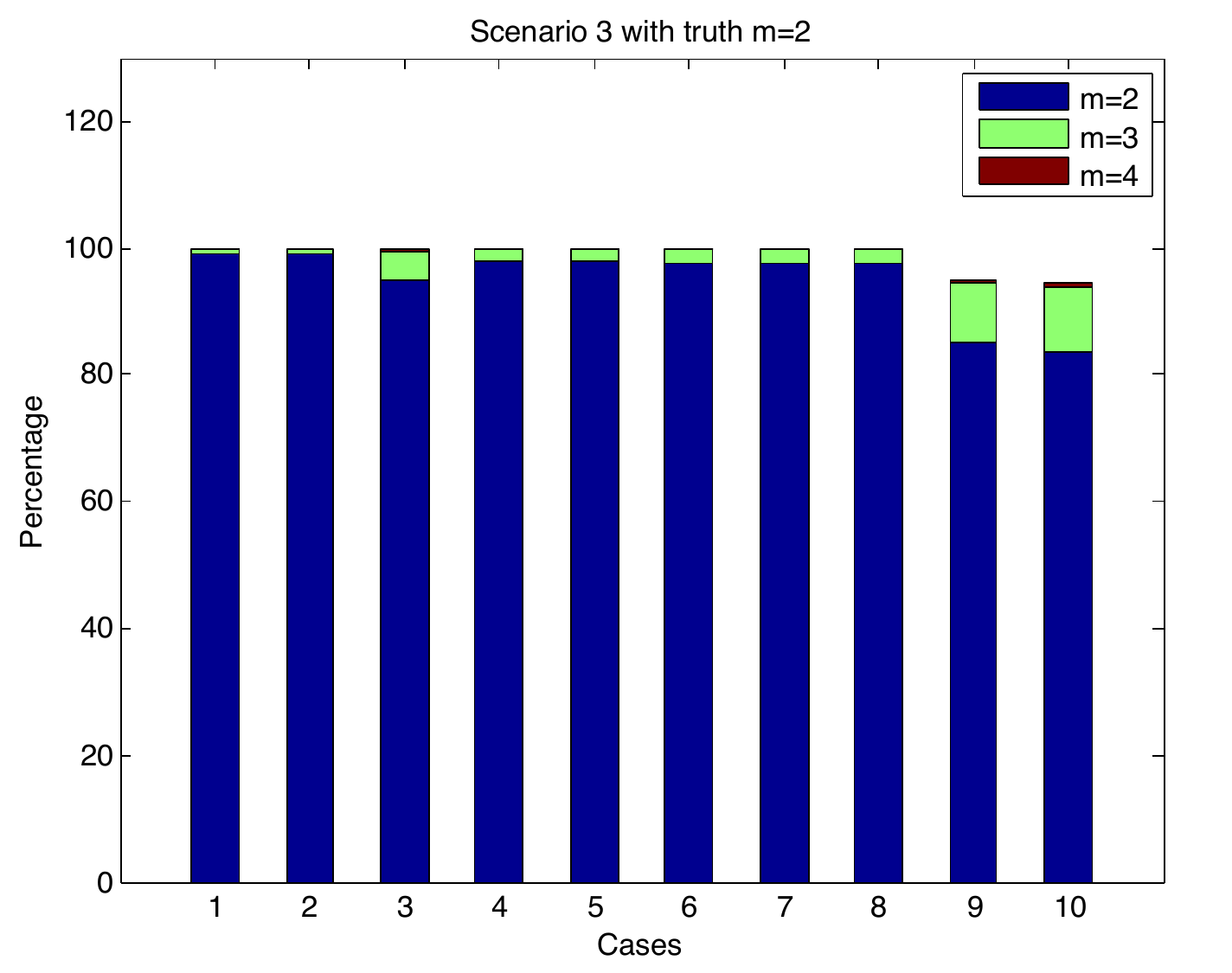}&\\
  Scenario 3&\\
\end{tabular} 
 \caption{The proportion of posterior estimates of $m$ in all cases for scenarios 1-3
   in the simulation. }
 \label{fig:m}
 \end{figure}

Figure \ref{fig:m} shows the proportion of posterior
estimates of the degree of polynomial $m$ in scenarios 1-3. Recall
that $m$ follows a discrete uniform prior taking values in the set
$\{0, 1,2,\ldots,M\};$ here we set $M=5$.  We can see that the
degree of the polynomial from the true model dominates the posterior
estimates.  In scenarios 4-5 there are no true $m$ values since the
true $\mu(t)$ is not a polynomial function. 

For each case, we computed the average posterior predictive
probability of response at a future time point $t_s$ over 100
patients, defined as
\\$q_s=\frac{1}{100}\sum_{j=1}^{100}Pr\{e_j(t_s)=1\mid \eb^K\},$ where
$Pr\{e_j(t_s)=1\mid \eb^K\}$ was calculated according to
\eqref{forecast} and \eqref{qhat}.  We used $t_s = 33,
34$ and $35$, since the lupus trial protocol plans to make forecasts
at these three future time points.  Also, we computed the posterior
mean DDR 
$$\hat{T}=\frac{1}{B}\sum_{b=1}^Bl\{t: \mu(t;
\bbeta^{(b)}, m^{(b)})>a_h\}.$$
 Table \ref{table:summary}  summarizes the results. 
 
 \begin{table}
\caption{\label{table:summary} True probability, empirical probability
  and average posterior predictive estimate (PPE) of response probability at future
  time point $t_{33}$, $t_{34}$ and $t_{35}$. The posterior mean DDR,  
  $\hat{T}$,  is also presented.}
  \centering
\begin{tabular}{llllll}
&   & $q_{33}$ & $q_{34}$ & $q_{35}$&$\hat{T}$\\
\hline
Scenario 1 & Truth &0.2130   & 0.2130  &  0.2130&0\\ 
               &  Empirical&0.28  &  0.22 &   0.25  &  8.87 \\ 
$K=32, a_h=0$               & PPE& 0.2212 &   0.2398  &  0.2491&0\\
$K=32, a_h=0.5$& PPE&0.2205   & 0.2389  &  0.2488&0\\
$K=29, a_h=0$ & PPE&0.2683   & 0.2637  &  0.2613&0\\
$K=29, a_h=0.5$&PPE&0.2712 &   0.2667 &   0.2640&0\\
$K=26, a_h=0$&PPE&   0.2734 &   0.2691  &  0.2663&0\\
$K=26, a_h=0.5$& PPE&0.2726 &   0.2687&    0.2665&0\\
$K=23, a_h=0$& PPE&  0.2445   & 0.2443  &  0.2442&0\\
$K=23, a_h=0.5$& PPE& 0.2450 &   0.2447  &  0.2446&0\\
$K=20, a_h=0$& PPE&    0.2377  &  0.2376  &  0.2376&0\\
$K=20, a_h=0.5$& PPE& 0.2399   & 0.2399   & 0.2399&0\\
\hline
Scenario 2 & Truth &0.6976   & 0.7113  &  0.7248&15\\ 
               &  Empirical&0.74  &  0.75 &   0.76  &  18.06 \\ 
$K=32, a_h=0$               & PPE& 0.7459   & 0.7568&    0.7717&18.807\\
$K=32, a_h=0.5$&PPE&0.7482   & 0.7604   & 0.7755&18.701\\
$K=29, a_h=0$ & PPE&0.7778 &  0.7920   & 0.8053&19.008\\
$K=29, a_h=0.5$& PPE&0.7800  &  0.7944 &   0.8075&19.142\\
$K=26, a_h=0$& PPE&   0.7684  &  0.7826 &   0.7960&18.762\\
$K=26, a_h=0.5$& PPE&0.7747  &  0.7878&    0.8010&18.896\\
$K=23, a_h=0$& PPE&  0.7906  &  0.8034 &   0.8156&18.892\\
$K=23, a_h=0.5$&PPE&0.7979  &  0.8106 &   0.8227&19.353\\
$K=20, a_h=0$& PPE& 0.7732  &  0.7852   & 0.7967&17.985\\
$K=20, a_h=0.5$& PPE& 0.7718  &  0.7840   & 0.7957&18.269\\
\hline
Scenario 3 & Truth & 0.3676  &  0.2557 &   0.1599&28.723\\ 
               &  Empirical&0.42  &  0.30 &   0.19  &  27.55 \\ 
$K=32, a_h=0$               & PPE& 0.4179  &  0.3140  &  0.2244&29.457\\
$K=32, a_h=0.5$& PPEn&0.4212  &  0.3201  &  0.2317&29.506\\
$K=29, a_h=0$ & PPE& 0.4763  &  0.3723  &  0.2731&29.893\\
$K=29, a_h=0.5$& PPE&0.4594 &   0.3538  &  0.2546&29.728\\
$K=26, a_h=0$& PPE& 0.6648   & 0.5795   & 0.4873&31.261\\
$K=26, a_h=0.5$& PPE&0.6465 &   0.5625  &  0.4742&30.962\\
$K=23, a_h=0$& PPE&0.4690   & 0.3814  &  0.2999&29.547\\
$K=23, a_h=0.5$& PPE&0.3744 &   0.2866  &  0.2105&28.762\\
$K=20, a_h=0$& PPE&0.7575  &  0.7160 &   0.6731&30.931\\
$K=20, a_h=0.5$& PPE& 0.7483  &  0.7068 &   0.6646&30.869\\
\end{tabular}
\end{table}

\begin{table}
\centering
\begin{tabular}{llllll}
&   & $q_{33}$ & $q_{34}$ & $q_{35}$&$\hat{T}$\\
\hline
Scenario 4 & Truth &0.2610   & 0.1349 &   0.0669&8.194\\
               &  Empirical&0.31&  0.18  &  0.12  &10.980\\
$K=32, a_h=0$               & PPE&   0.3118 &   0.1660   & 0.0823&9.985\\
$K=32, a_h=0.5$ &PPE&  0.3123  &  0.1664   & 0.0825 &9.988\\
$K=29, a_h=0$       & PPE&  0.2458   & 0.1228  &  0.0614 &9.687\\
$K=29, a_h=0.5$  & PPE&   0.2458 &   0.1228  &  0.0614 &9.687\\
$K=26, a_h=0$  & PPE&   0.2640   & 0.1366 &   0.0683 &9.095\\
$K=26, a_h=0.5$ & PPE&    0.2687  &  0.1400 &   0.0699 &9.093\\
$K=23, a_h=0$ & PPE&    0.2528 &   0.1314  &  0.0677 &9.143\\
$K=23, a_h=0.5$  & PPE&  0.2522   & 0.1310   & 0.0675 &9.143\\
$K=20, a_h=0$ & PPE&  0.2641  &  0.1387   & 0.0710&9.230\\
$K=20, a_h=0.5$ & PPE&   0.2632  &  0.1378   & 0.0705&9.231\\
\hline
Scenario 5 & Truth &  0.2104 &   0.1720  &  0.1599&20.00\\ 
               &  Empirical&0.23  &  0.22 &   0.19  & 19.54 \\ 
$K=32, a_h=0$               &PPE& 0.1544  &  0.0679 &   0.0181&23.62\\
$K=32, a_h=0.5$& PPE&0.1447  &  0.0711   & 0.0287&23.06\\
$K=29, a_h=0$ & PPE&0.0068  &  0.0002 &   0.0000&19.52\\
$K=29, a_h=0.5$& PPE& 0.0103  &  0.0003  &  0.0000&19.66\\
$K=26, a_h=0$& PPE&0 &  0  &  0&17.70\\
$K=26, a_h=0.5$& PPE&0 &  0  &  0&17.59\\
$K=23, a_h=0$& PPE&0.0035   & 0.0028  &  0.0024&16.66\\
$K=23, a_h=0.5$& PPE&0.0017  &  0.0012 &   0.0011&16.87\\
$K=20, a_h=0$& PPE&1 &  1  &  1&25.79\\
$K=20, a_h=0.5$& PPE&0.9960 &   0.9950  &  0.9950&25.73\\
\hline
\end{tabular}
\end{table}

\begin{figure}[h!]
\includegraphics[width=0.5\textwidth]{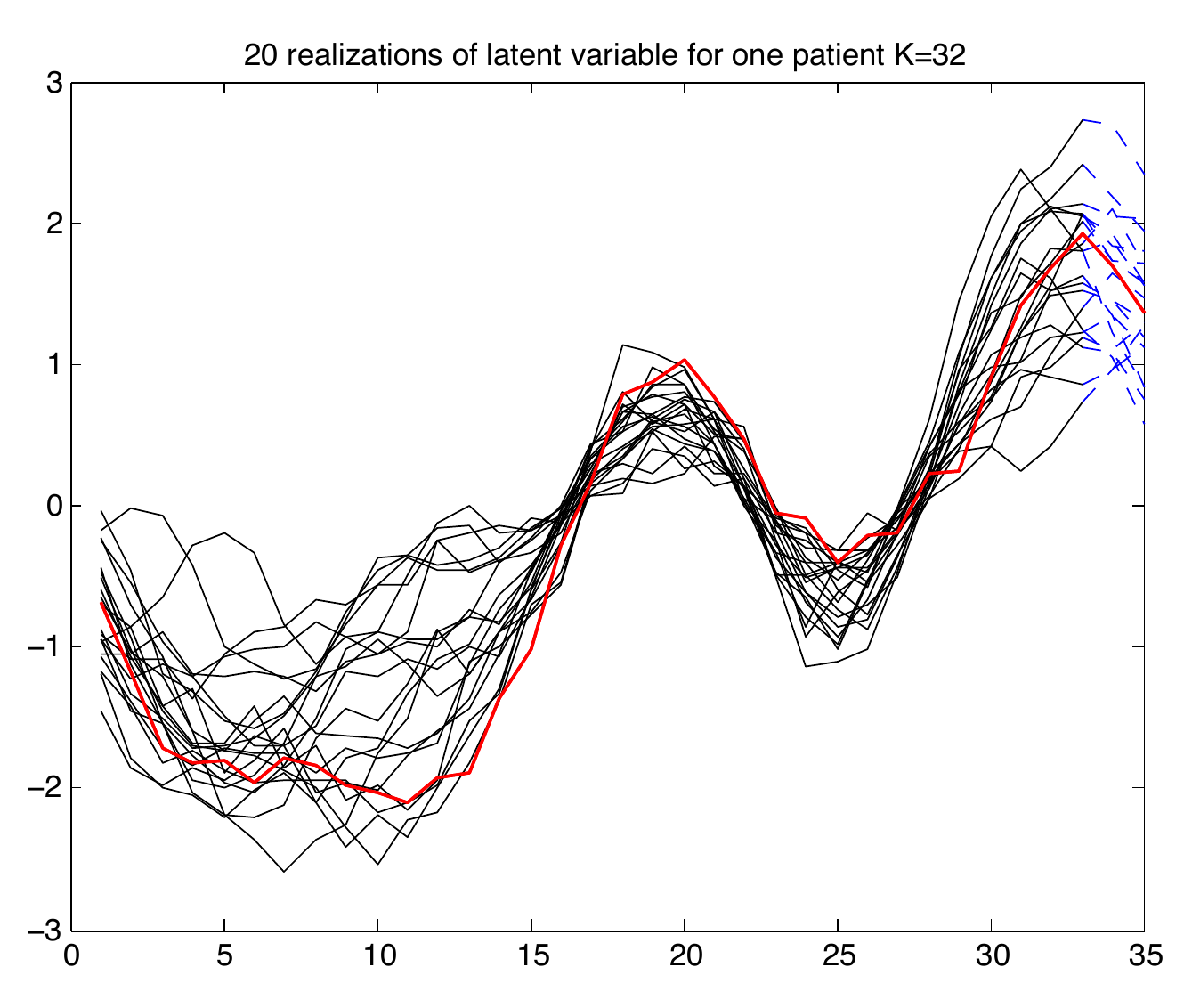}
\includegraphics[width=0.5\textwidth]{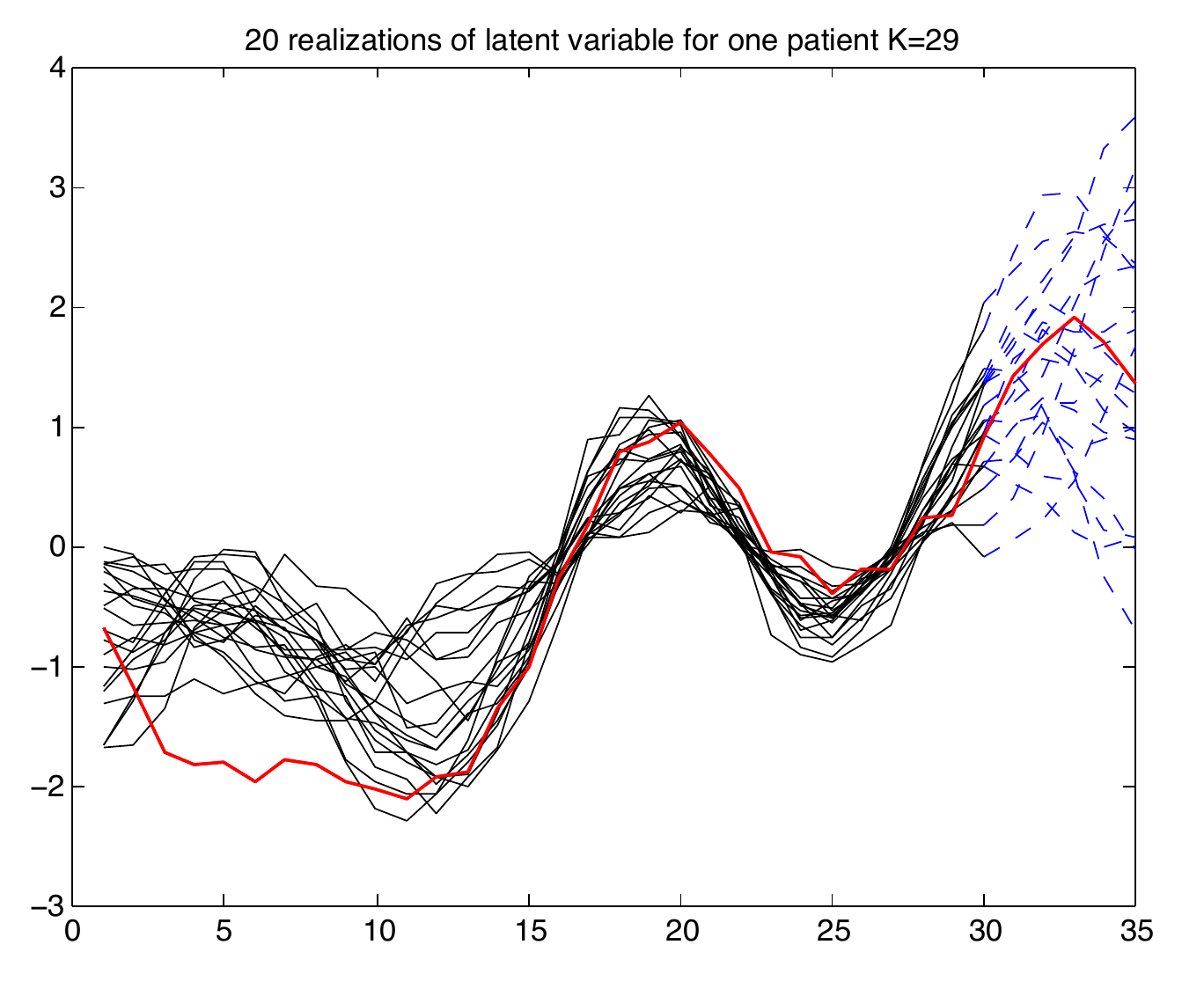}
\caption{A set of 20 realizations of latent variables for two cases in
  scenario 2. The left panel presents the latent variable values for
  one patient with the first 32 time points observed. The right panel
  presents one patient with the first 29 time points observed. The red
  solid line represents the truth. Black solid lines represent 20
  realizations at observed time points, and blue dashed lines represent predicted values at future time points.}
\label{fig:reali2}
\end{figure}

\begin{figure}[h!]
\includegraphics[width=0.5\textwidth]{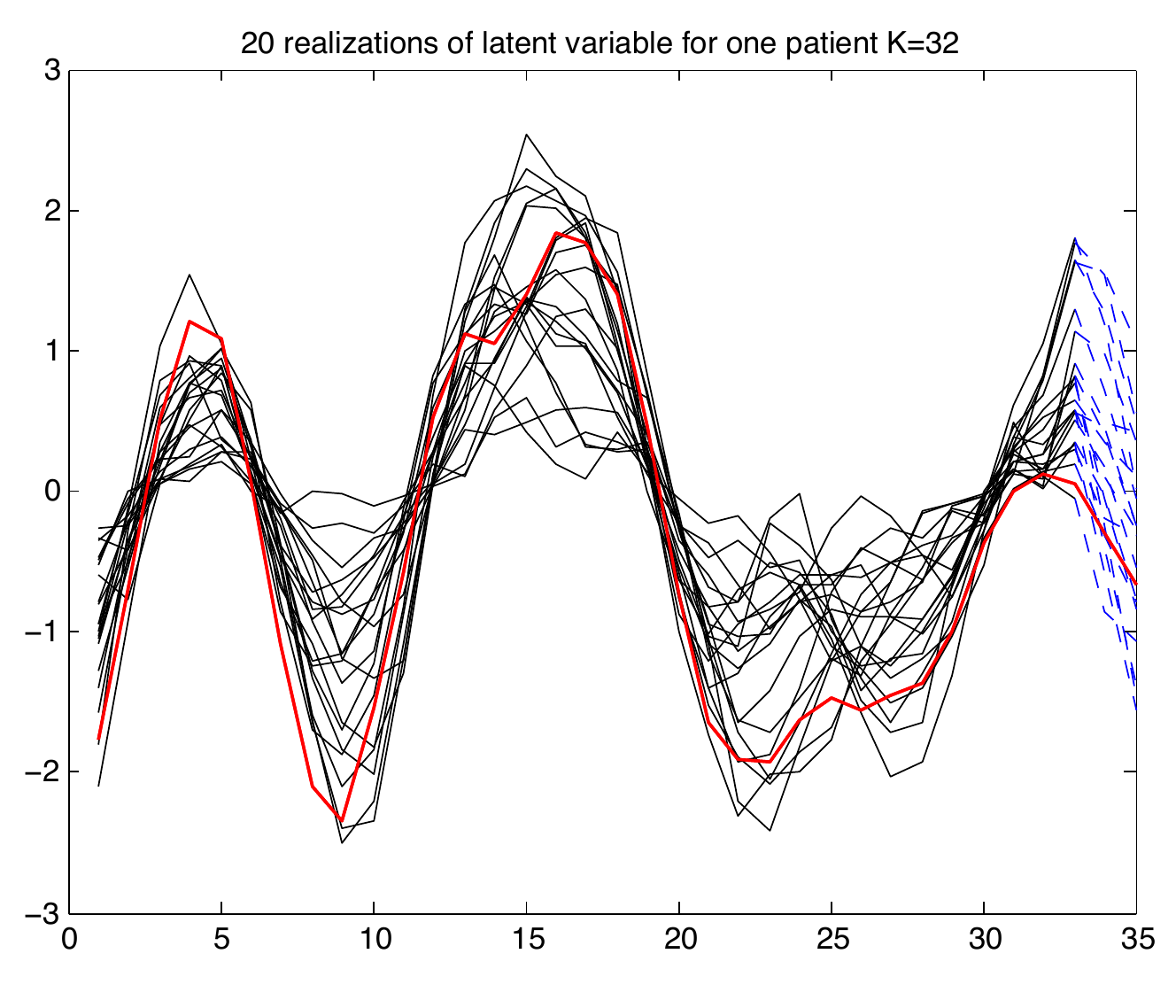}
\includegraphics[width=0.5\textwidth]{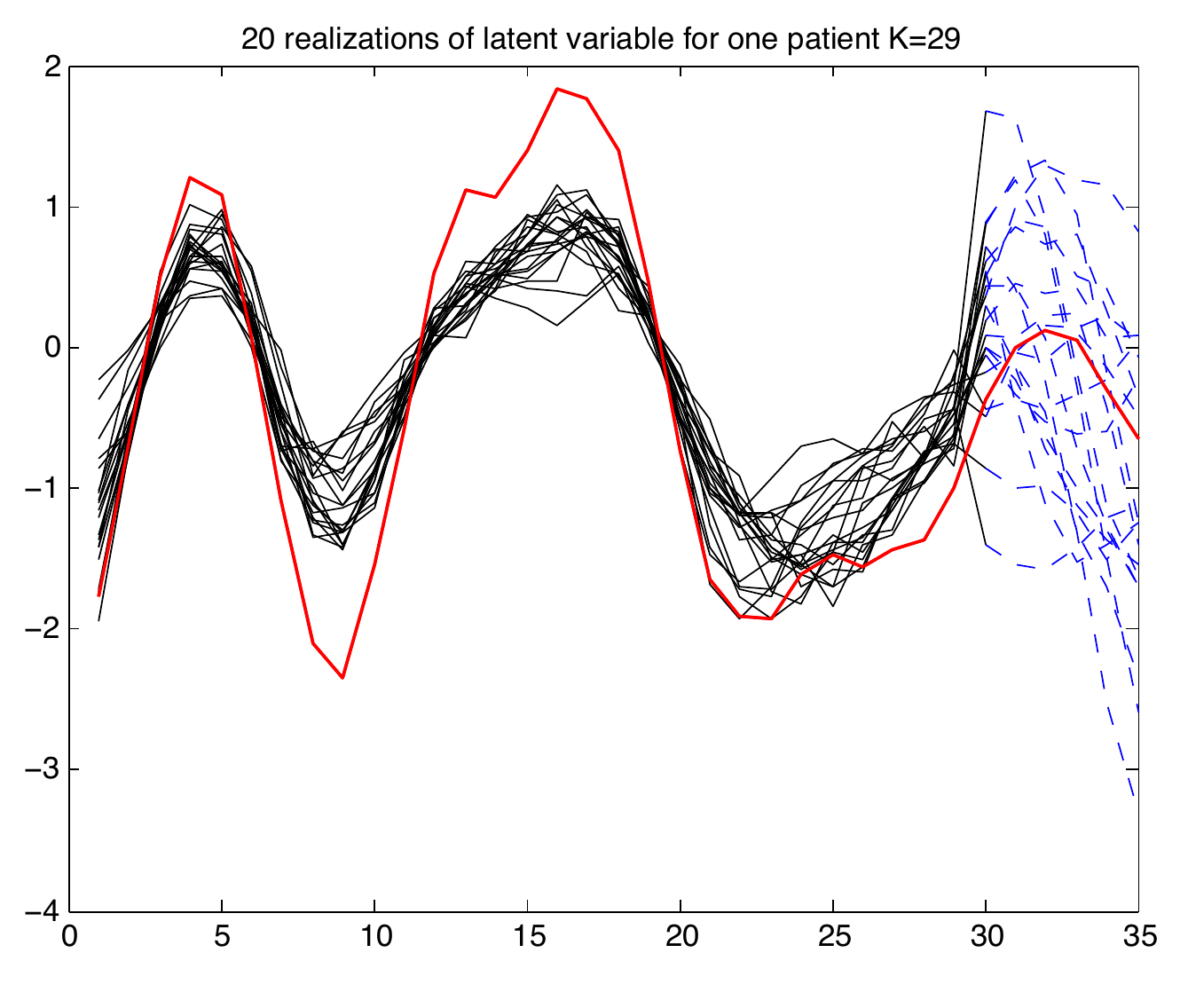}
\caption{A set of 20 realizations of latent variables for two cases in
  scenario 4. The left panel presents the latent variable values for
  one patient with the first 32 time points observed. The right panel
  presents one patient with the first 29 time points observed. The red
  solid line represents the truth. Black solid lines represent 20
  realizations at observed time points, and blue dashed lines represent predicted values at future time points.}
\label{fig:reali4}
\end{figure}

After obtaining the posterior imputation of the latent variables ${\bm
  a}^K$ and posterior predictive values of ${\bm a}^{(35-K)}$ for the
new time points, we plotted 20 realizations of the latent variables
for one randomly selected patient in case $(K=32, a_h=0)$ for scenario
2 and case
$(K=29, a_h=0)$ for scenario 4 as shown in Figures
\ref{fig:reali2} and \ref{fig:reali4}, respectively.  The left panel presents the latent variable values given the first 32 observed time points and the right panel presents those with the first 29 time points observed. It can be seen that our model fits the simulated data well in both panels. 

In scenario 5, we generated simulated data using the model with a
trigonometric mean function and covariance function (\ref{coveq})
without trigonometricity, but
fitted the model with the polynomial mean function and the
trigonometric covariance function
(\ref{eq:period}) to the simulated data. As seen in Table \ref{table:summary},
the predictions are poor. This implies that the model choice is
important in terms of the placement of the trigonometric functions, in the mean or covariance of the GP. 

Next we checked the robustness of the choice of $a_h$. 
As shown in Table \ref{table:summary}, 
 the
 estimations remained consistently well  when we varied the values of $a_h$ ($a_h=0$ and $a_h=0.5$). Thus, the model is robust to choice of $a_h$.

Finally we examined the impact of different $K$'s on the forecast.
From
Table \ref{table:summary} and Figure \ref{fig:m}, we can see that the fewer
the observed time points, i.e. the smaller the value of $K$, 
the worse the prediction. For example, in scenario 3, when $K=32$ and
$a_h=0$, the degree of polynomial of the true model is successfully
recovered $99\%$ of times;  but when $K=20$ and
$a_h=0$, this percentage drops to $85\%$. Also, the true
$q_{33}=0.3676$, while the posterior estimate of $q_{33}$ is 0.4179
when $(K=32, a_h=0)$ and 0.7575 when $(K=20, a_h=0)$. 
This also echoes
the consistency results in Section \ref{sec:post} to be shown next;  that is, the larger the number of time
points observed, the better the estimation. 

Next in the lupus trial example,  one concern is how to choose the
proper time point to start the forecast, which also depends on the
drug effects. Based on previous clinical experiments and prior knowledge, we
can make assumptions about the mean of the the drug effects and the
start time point can be determined by simulation and
calibration. Examining the simulation results observed so far, we
decided to use $K=23$ as the starting time for forecast. 

\section{Trial Example}
\label{sec:trial}
Using the lupus trial as an example, we simulated a large number
of clinical trials on computer and applied the proposed monitoring rules
to examine the operating characteristics of our method. 
According to the original trial setup, we set
the duration of the
simulated trials at $35$ weeks and the maximum number of
patients to be 200, equally
randomized between the standard arm and experimental
arm. We assumed that patients were recruited over time and the
number of patients enrolled weekly for each arm was  a random number following
a discrete uniform in $\{2, 3, 4\}$. Patients were observed and
followed each week, when their responses $e$'s were recorded. The
experimental treatment (arm 2) is considered more effective than the
control (arm 1) if the DDR in \eqref{eq:r-time} for the
treatment arm $T_2$ is at least 2 weeks longer than that of the
control arm $T_1$, i.e., the minimum difference between the two arms must be $\delta=2$. 

We considered six scenarios and simulated 100 trials for each
scenario. The binary response
outcomes were generated over time using the following scheme. We first specified 
true  mean functions (see Table \ref{table:setup} for different
$\mu_i(t)$)  
and the
true covariance function (\ref{eq:period}) with $\theta_1=1$,
$\theta_2=3.5$ and $r=2$. We then generated the LGP $\ba(t_k)$ and
$\eb(t_k)$ at 35 weeks of follow up, $k=1, 2, \ldots,
35$.

\begin{figure}[h!]
\centering
\includegraphics[width=.32\textwidth, height=60mm]{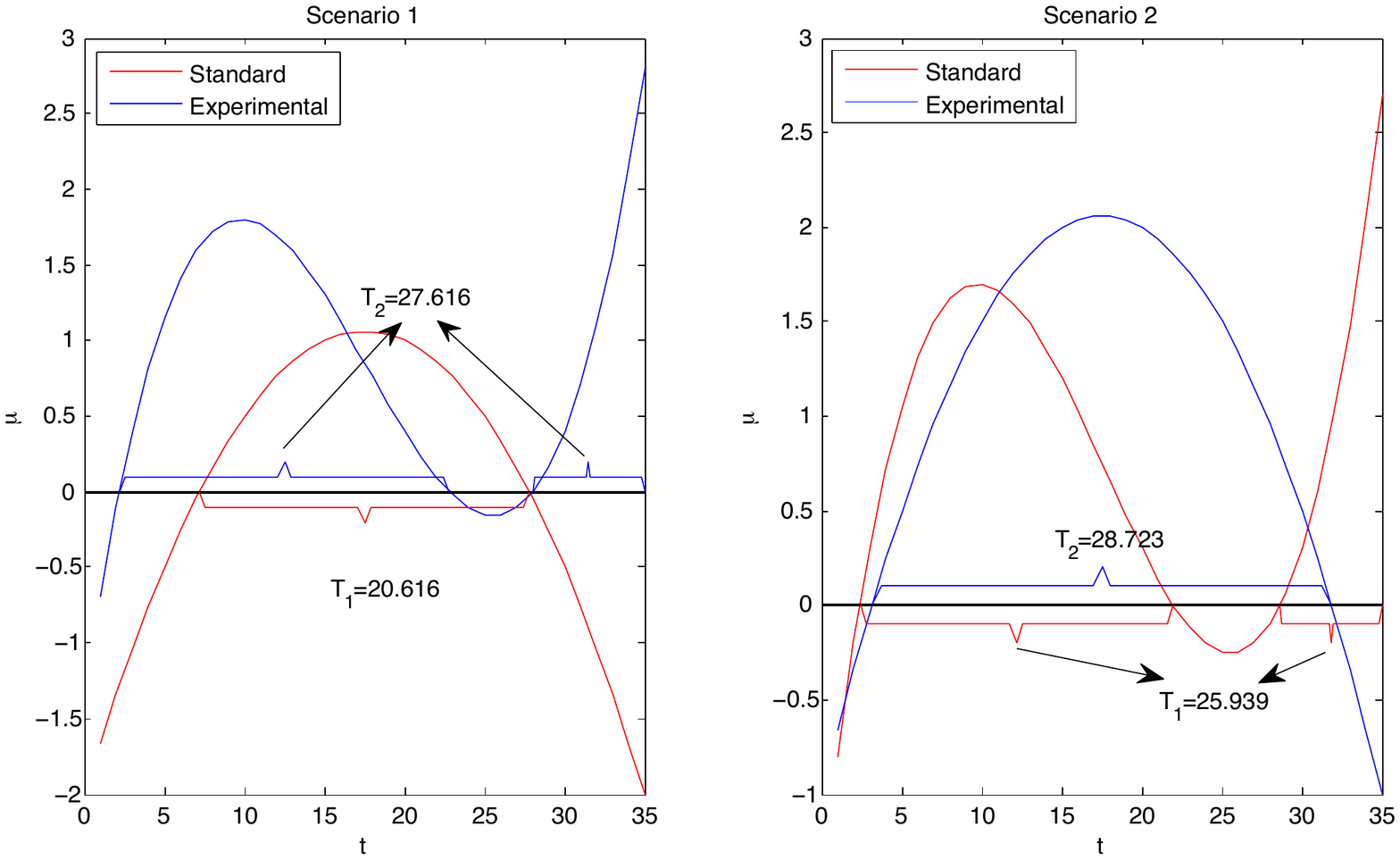}
\includegraphics[width=.32\textwidth, height=60mm]{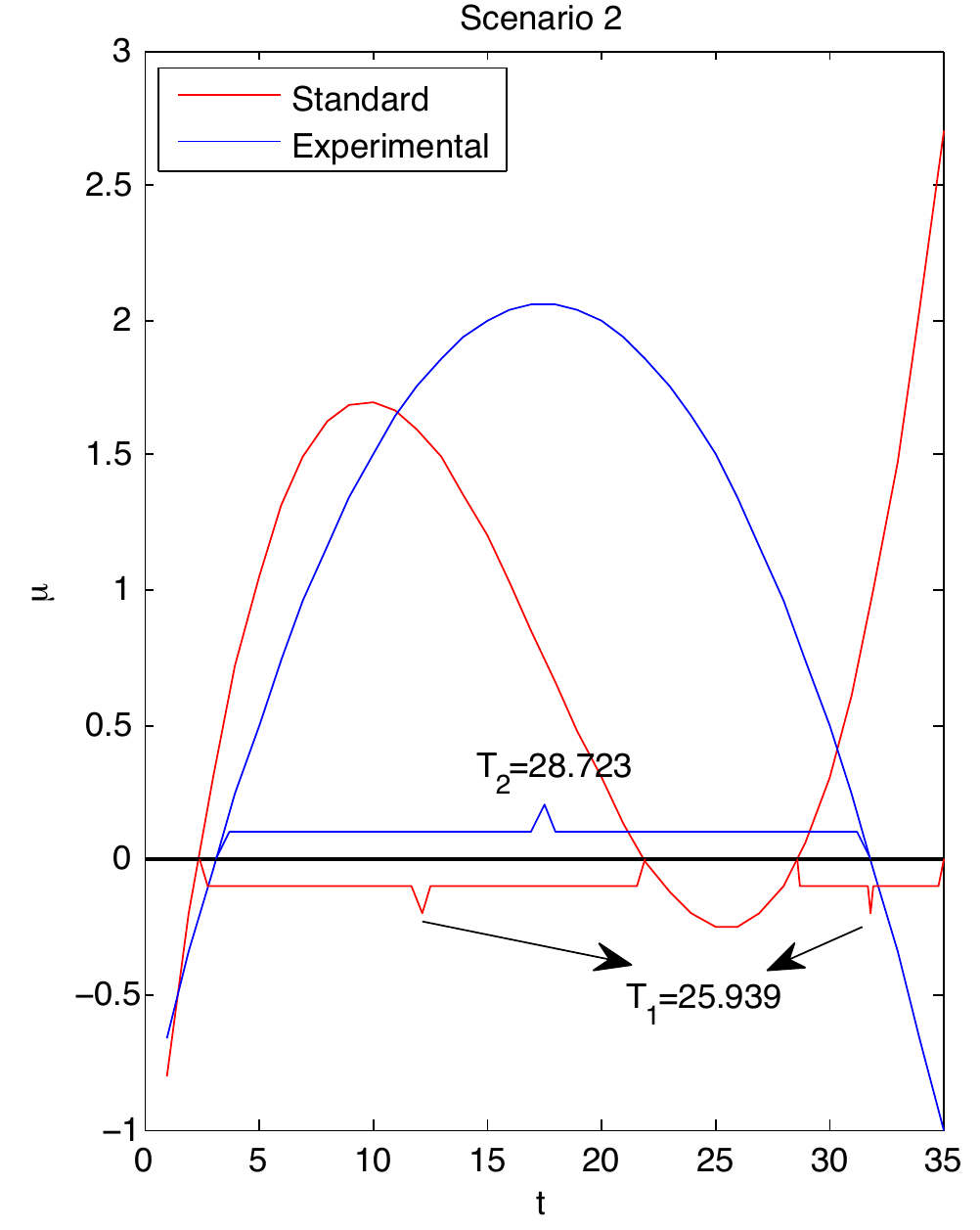}
\includegraphics[width=.32\textwidth,height=61mm]{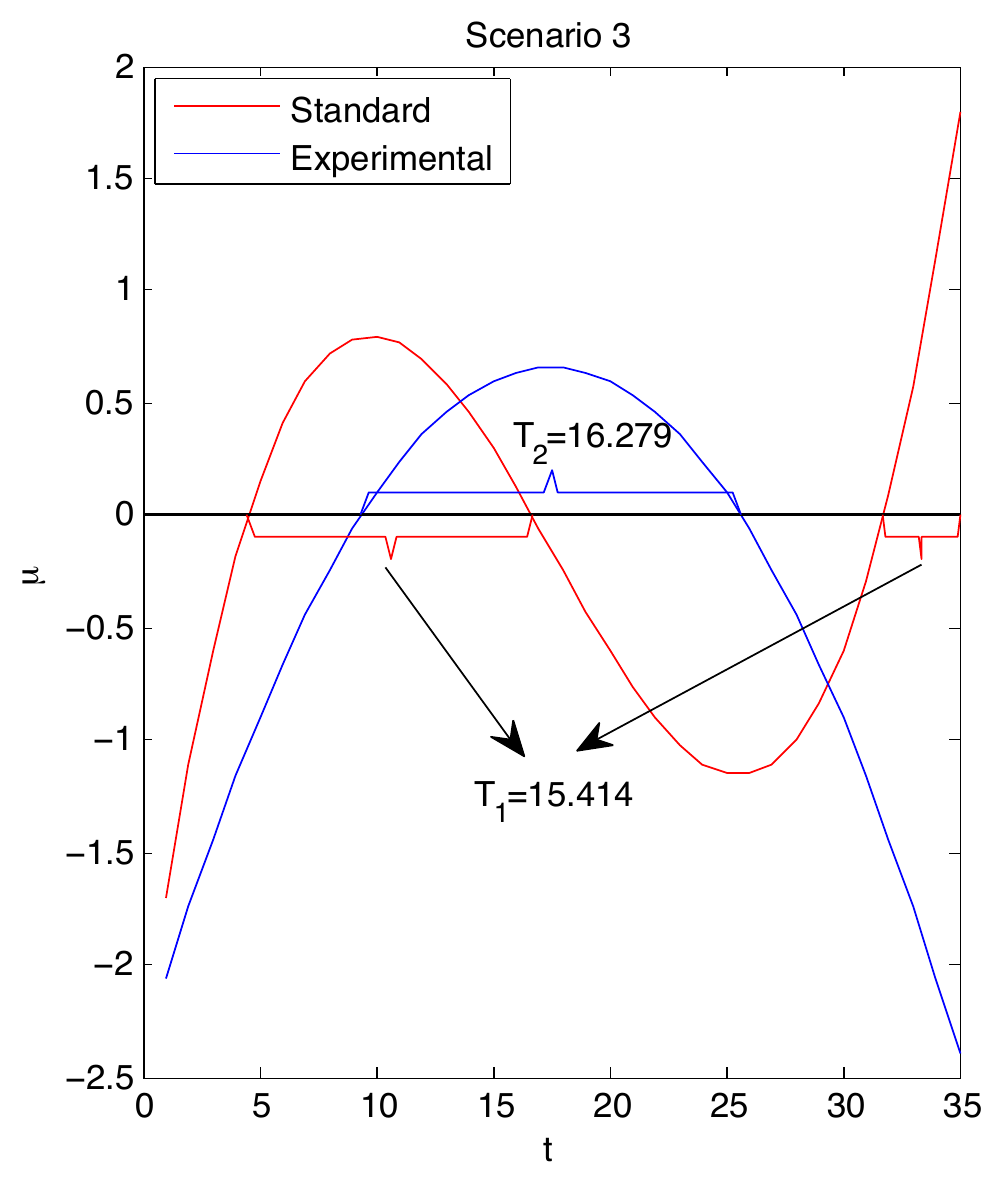}
\includegraphics[width=.32\textwidth, height=60mm]{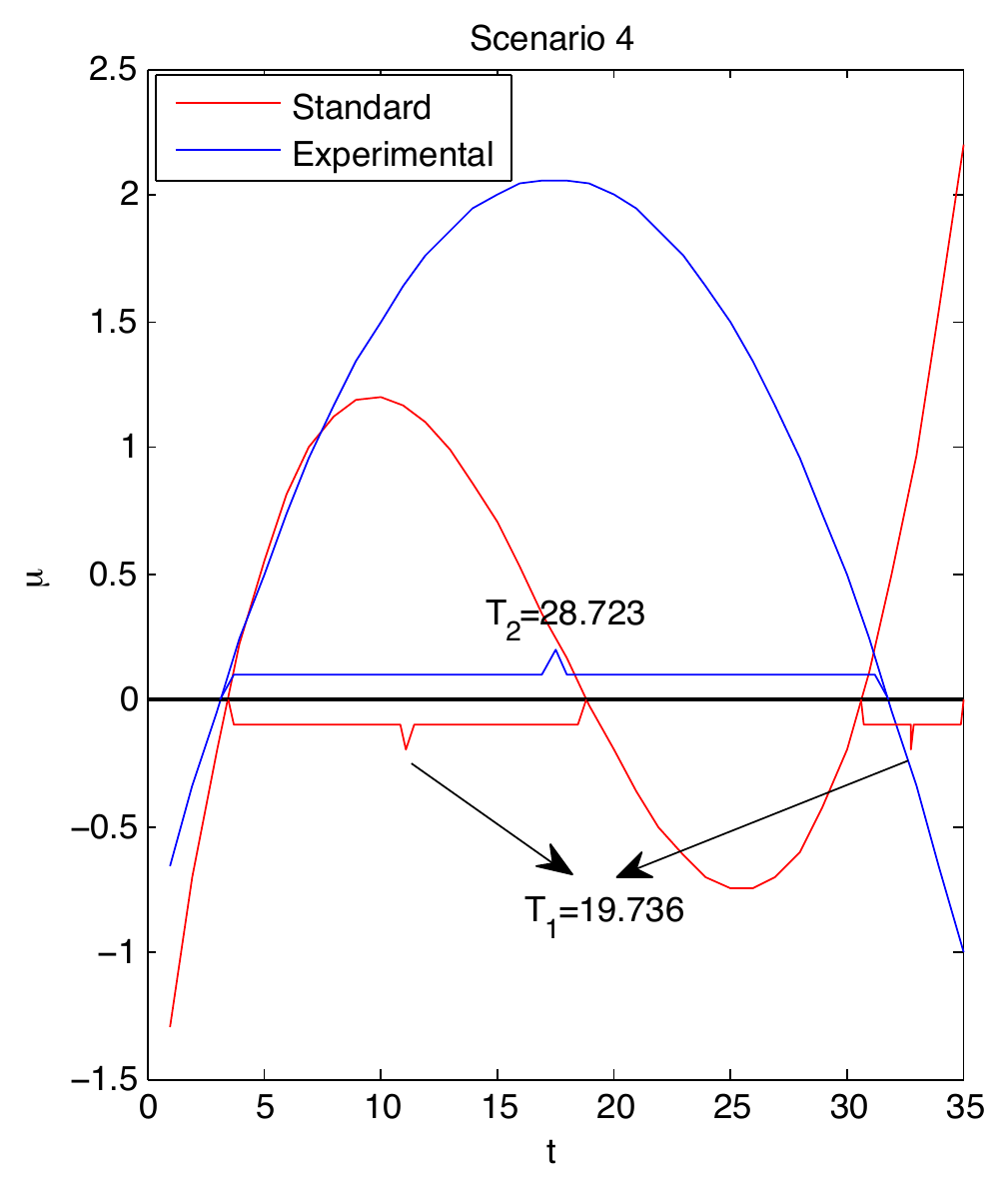}
\includegraphics[width=.32\textwidth, height=60mm]{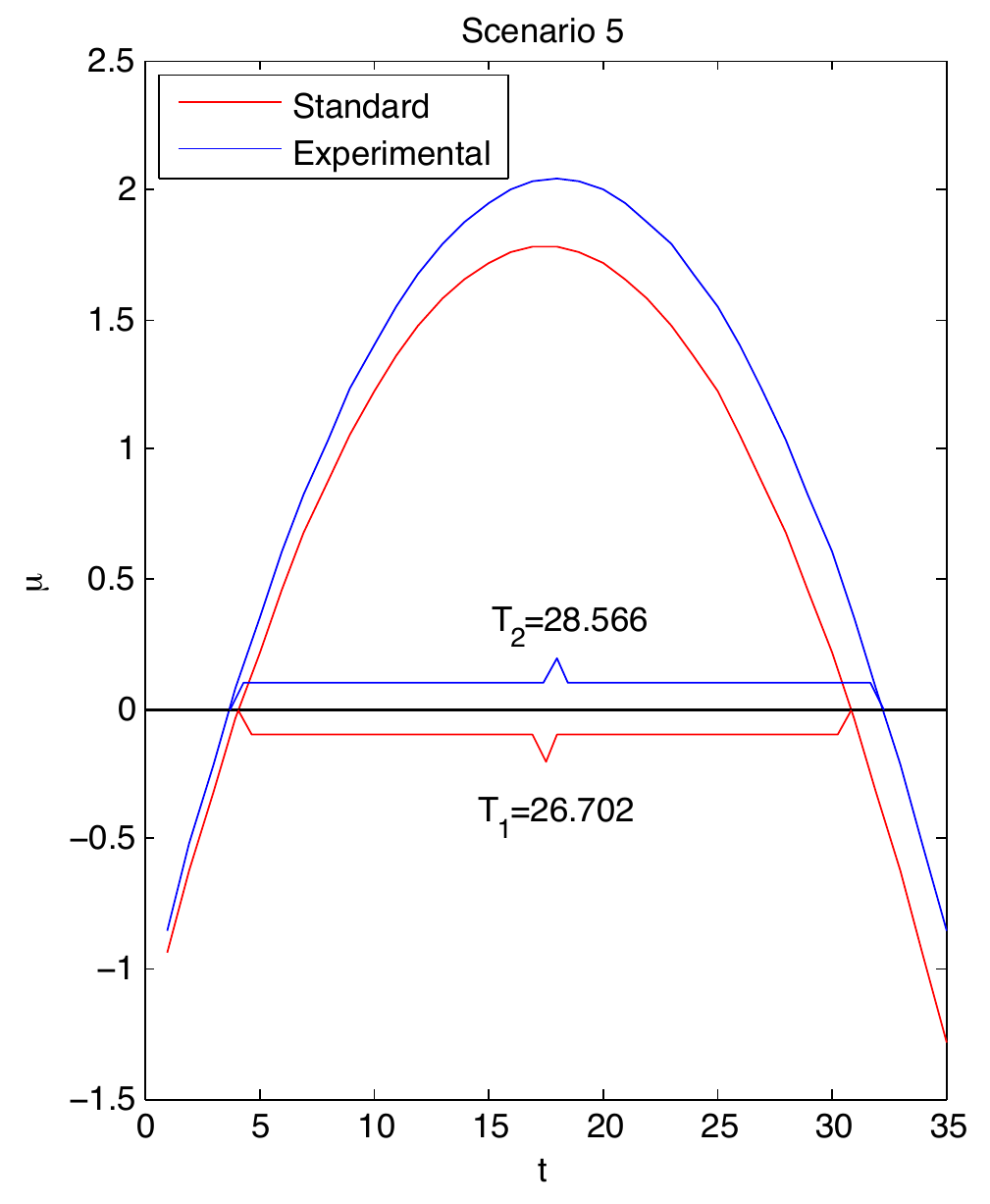}
\includegraphics[width=.32\textwidth,height=60mm]{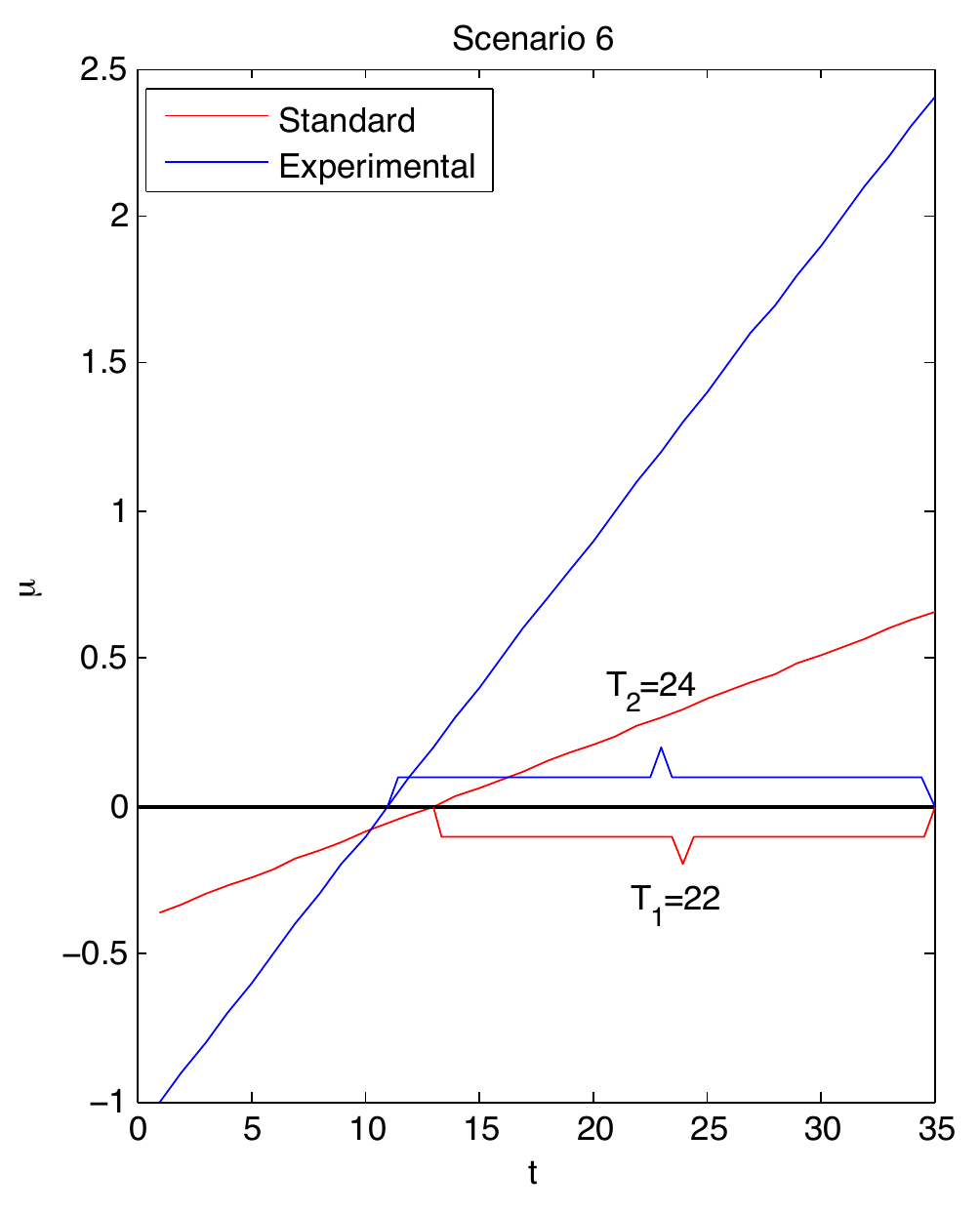}
\caption{The true DDR $T_i$ and the true mean functions $\mu_i(t)$ in the control 
  and experimental arms in all six scenarios for the lupus
  trial simulation. The red line represents the true mean in the control 
  arm, while the blue line represents the true mean in the experimental arm. The black horizontal line represents the threshold $a_h=0$.}
\label{fig:truth}
\end{figure}

For all six scenarios, the true DDR $T_i$ according to
the true
$\mu_i(t)$ were calculated, and the
configurations including $m_i$, $\bbeta_i$ and
$T_i$, $i=1, 2$ are displayed in Table
\ref{table:setup}.  Figure \ref{fig:truth} shows the true mean
functions of the standard and experimental arms in all six
scenarios, in which we also mark the true $T_i$ values. The arm with a larger $T_i$ value is more effective since it leads
to a longer DDR. For example, in scenario 1,
$T_1=20.616$ and $T_2=27.616$, so arm 2, the experimental treatment, is
better. 

\begin{table}
\caption{\label{table:setup} True Values of $m_i$, $\bbeta_i$ and  $T_i$, $i=1, 2$ for all six scenarios in the simulation.}
\centering
\begin{tabular}{lllllllr}
\hline
  &Group & $m_i$&$\bbeta_i$&$T_i$ \\
\hline
Scenario 1 & Standard &2& (-2, 3.5, -1)&20.6 \\ 
                     & Experimental  &3& (-1.4, 7.5, -5.3, 1)&27.6\\ 
                     \hline
Scenario 2 & Standard &3&(-1.5, 7.5, -5.3, 1)&25.9\\ 
                     & Experimental&2&(-1, 3.5, -1)&28.7\\ 
\hline
Scenario 3 & Standard &3&(-2.4, 7.5, -5.3, 1)&15.4\\ 
                     & Experimental&2&(-2.4, 3.5, -1)&16.3\\ 
                             \hline
Scenario 4 & Standard &3&(-2, 7.5, -5.3, 1)&19.7\\ 
                     & Experimental&2&(-1, 3.5, -1)&28.7\\ 
\hline 
Scenario 5 & Standard &3&(-1.28, 3.5, -1)&26.7\\ 
                     & Experimental&3&(-1.2, 3.6, -1)&28.6\\ 
\hline 
Scenario 6 & Standard &2&(-0.39, 0.3)&22.0\\ 
                     & Experimental&2&(-1.1, 1)&24.0\\ 
\hline 
\end{tabular}
\end{table}

We fitted the proposed model (\ref{eq:joint}) 
to the simulated data and assumed vague priors ${\bm \beta}\sim \text{Normal}({\bm 0}, 10^2{\bf I})$,  $\theta_1, \theta_2, r \sim \text{Normal} (0, 10^2)$.
The trial monitoring started at week $K=23$ and was based on the proposed monitoring rules
with $\delta=2$. 
Table \ref{table:trial} summarizes the operating characteristics of the LGP method for all six scenarios, including the average duration of trials, the maximum duration of trials, the average number of patients studied and stopping probabilities.

\begin{table}
\caption{\label{table:trial} Results for the simulations based on the lupus trial. AD: average trial duration (weeks); MD: maximum duration (weeks); AP: average number of patients in each group.}
\begin{tabular}{lllllllr}
\hline
  &Group & $T_i$&AD&MD& AP &Stopping probabilities\\
\hline
Scenario 1 & Standard & 20.6&24.9&29&76.1& 97\%(superiority) 
\\ 
                     & Experimental  &27.6\\ 
                     \hline
Scenario 2 & Standard &25.9& 29.72&35&87.43&52\%(superiority) 
\\ 
                     & Experimental&28.7\\ 
\hline
Scenario 3 & Standard  & 15.4& 29.06&35&87.68&60\%(futility) 
\\ 
                     & Experimental  &16.3\\ 
                              \hline
Scenario 4 & Standard &19.7&23.31&26&70.65& 100\%(superiority)\\ 
                     & Experimental&28.7\\ 
\hline
Scenario 5 & Standard &26.7&29.11&35&86.91&25\%(superiority) and 30\%(futility)\\ 
                     & Experimental&28.6\\ 
\hline 
Scenario 6 & Standard &22.0&30.90&35&91.59&16\%(superiority) and 19\%(futility)\\ 
                     & Experimental&24.0\\ 
\hline 
\end{tabular}
\end{table}

 In scenario 1, $m_1=2$ and $m_2=3$, and from Figure
 \ref{fig:truth}, patients in the experimental arm relapse
 more often than patients in the control arm. However, the
 experimental arm is still preferred since it has longer DDR 
 as
 $T_2=27.616>(20.616+2)=(T_1+\delta)$. Thus, scenario 1 explored the ability of the
 proposed LGP method to stop if the experimental arm has a longer DDR but
 more frequent relapses. The average trial duration was 24.9 weeks
 and the maximum trial duration was 29 weeks, which indicated that
 the trial stopped quickly after it was monitored at week 23.  The
 average number of patients was 76.1 per group and the maximum was
 100 patients. Among 100 simulated trials, 97 were stopped early due to superiority. 
 
In scenario 2, $m_1=3, m_2=2$ and $T_1=25.939$, $T_2=28.723$. Here
$T_2 - T_1 \approx 2.8$ which is close to $\delta=2$.  
 Among the 100 trials simulated, 52 stopped early due to
 superiority. 
 In this scenario, the average trial duration was 29.72 weeks and the maximum trial duration was 35 weeks. The average number of patients was 87.43 per group. 

In scenario 3, $m_1=3, m_2=2$ and $T_1=15.414$, $T_2=16.279$. Among
the 100 simulated trials, 60 stopped early due to
futility as expected. 
 In this scenario, the average trial duration was 29.06 weeks. The
 average number of patients was 87.68 per group.  In scenario 4,
 $m_1=3, m_2=2$ and $T_1=19.736$, $T_2=28.723$, implying that arm 2 was
 much better than arm 1. All 100 trials stopped early due to
 superiority of arm 2. The average trial duration was 23.31 weeks and the maximum trial duration was 26 weeks. The average number of patients was 70.65 per group.

In the last two scenarios, different arms had the same $m_i$ value. In
scenario 5, $m_1=m_2=2$. The average trial duration was 29.11 weeks,
and the average number of patients was 86.91 per group. Among the 100
trials simulated, 25 stopped due to superiority, 30 stopped due to
futility, and 45 did not stop early. 
In scenario 6, 
$m_1=m_2=1$ and $T_1=22$, $T_2=24$. The DDR in
arm 2 was exactly 2 weeks longer than in arm 1, making it  
difficult to stop early. Among 100 simulated trials, 65 didn't stop
early, 16 stopped early due to superiority, and 19  stopped early due to futility. The average trial duration was 30.9 weeks and the maximum trial duration was 35 weeks. The average number of patients was 91.59 per group.

In summary, the proposed LGP model and trial monitoring rules
exhibited desirable operating characteristics in all six scenarios.

\section{Discussion}
\label{sec:dis}
We have proposed a Bayesian LGP model for monitoring 
clinical trials with binary repeated outcomes.  
Through
the posterior estimates, we can predict the probability of efficacy
response for each patient at future time points. The proposed trial
monitoring rules allow for early termination of a trial if one arm is
considered more effective than the other. The Bayesian paradigm works very well and yields desirable results in our simulation studies. 

Although our  current models are set up for binary outcomes,
they can be easily extended to other applications with 
ordinal or categorical outcomes. Furthermore, the LGP provides a
framework for the inclusion of patient- or group-specific covariates such as patients' weights, ages, etc.,
which can be easily implemented by expanding the columns of the design
matrix $X_{ij}$ in \eqref{latent}. This will be a future direction
of our research. 

\section*{Acknowledgements}
Yuan Ji's research is partly supported by NIH grant R01 CA132897.

\section*{APPENDIX A: MCMC Details}
\subsection*{Joint Distributions}

\begin{eqnarray*}
P({\bm{e}}^K, {\bm{a}}^K, \bmm, {\bm \beta},  \Theta) 
&=& P({\bm{e}}^K\mid{\bm{a}}^K)P({\bm{a}}^K\mid {\bm \beta},  \bmm,  \Theta)P({\bm \beta}\mid \bmm)P(\bmm)P(\Theta) \nonumber\\
&=&\Bigg[\prod_{i=1}^{2}\bigg[ \prod_{j=1}^{N_i}\bigg\{\prod_{k=1}^{K_{ij}}\Big\{P(e_{ijk}\mid a_{ijk})=I(a_{ijk}>a_h)I(e_{ijk}=1)\nonumber\\
&&+I(a_{ijk}<=a_h)I(e_{ijk}=0)\Big\}\bigg\} \nonumber\\
&&\times (2\pi)^{-\frac{K_{ij}}{2}} |\bC(\bt_{ij})|^{-\frac{1}{2}}\nonumber\\
&&\times\exp\Big\{-\frac{1}{2}({\bm{a}}_{ij}^K-\bX_{ij}{\bm \beta}_i)'\bC(\bt_{ij})^{-1}({\bm{a}}_{ij}^K-\bX_{ij}{\bm \beta}_i)\Big\}\bigg]\nonumber\\
&&\times (2\pi)^{-\frac{m_i+1}{2}}|\sigma_0^2{\bf I}_{m_i+1}|^{-\frac{1}{2}}\nonumber\\
&&\times\exp\Big\{-\frac{1}{2}({\bm \beta}_i-{\bm \mu}_0^{m_i+1})'(\sigma_0^2{\bf I}_{m_i+1})^{-1}({\bm \beta}_i-{\bm \mu}_0^{m_i+1})\Big\}\Bigg] \nonumber\\
&&\times P(\bmm)P(\Theta) \nonumber\\
\end{eqnarray*}

where ${\bm{a}}^K_{ij}=(a_{ij1}, \dots, a_{ijK_{ij}})', {\bm \beta}_i=(\beta_{i0}, \beta_{i1}, \dots, \beta_{i,m_i})'$, ${\bf I}_{m_i+1}$ is an $(m_i+1)\times (m_i+1)$ dimension identity matrix, and

\[
\bX_{ij} =
 \begin{pmatrix}
  1 & t_1&\cdots &t_1^{m_i} \\
  1 & t_2&\cdots &t_2^{m_i}  \\
   \vdots  & \vdots &\ddots &\vdots  \\
  1 & t_{K_{ij}} &\cdots &t_{K_{ij}}^{m_i}  \\
 \end{pmatrix}.
\]

\subsection*{Full Conditional} 

\subsubsection*{Algorithm for simulating ${\bm{a}}_{ij}^K$}
\begin{eqnarray*}
P({\bm{a}}_{ij}^K\mid{\bm{e}}^K, {\bm \beta}, \bmm, \Theta)
&\propto&\Big\{\prod_{k=1}^{K_{ij}}P(e_{ijk}\mid a_{ijk})=I(a_{ijk}>a_h)I(e_{ijk}=1)\nonumber\\
&&+I(a_{ijk}<=a_h)I(e_{ijk}=0)\Big\} \nonumber\\
&&\times (2\pi)^{-\frac{K_{ij}}{2}} |\bC(\bt_{ij})|^{-\frac{1}{2}}\nonumber\\
&&\times\exp\Big\{-\frac{1}{2}({\bm{a}}^K_{ij}-\bX_{ij}{\bm \beta}_i)'\bC(\bt_{ij})^{-1}({\bm{a}}^K_{ij}-\bX_{ij}{\bm \beta}_i)\Big\}\nonumber\\
\end{eqnarray*}

To sample this truncated multivariate normal distribution, we use the method of \cite{geweke1991efficient} to compose a cycle of Gibbs steps through the components of ${\bm{a}}_{ij}$.


To draw the samples from multivariate normal distribution subject to linear inequality restrictions: 
$$x \sim N(\mu, \Sigma). \ \ \ \ \ \ \ \ a<Dx<b,$$
where $x$ is $n-$dimensional vector, $a$ and $b$ are $m-$dimensional vectors, and $D$ is $m\times n$ matrix imposing linear inequality restrictions. 

This is equivalent to:
\begin{equation}
z \sim N(0,T), \ \ \ \ \ \ \alpha<z<\beta.
\label{A1}
\end{equation}

where 
$$T=D\Sigma D', \ \, \alpha=a-D\mu, \ \, \beta=b-D\mu,$$
and we then take $x=\mu+D^{-1}z$.

Suppose that in the non-truncated distribution $N(0, T)$,

$$E[z_i\mid z_1, \dots, z_{i-1}, z_{i+1}, \dots, z_n]=\sum_{j\neq i}c_{ij}z_j.$$

Then in the truncated normal distribution of (\ref{A1}), the distribution of $z_i$ conditional on $\{z_1, \dots, z_{i-1}, z_{i+1}, \dots, z_n\}$ has the construction

$$z_i=\sum_{j\neq i}c_{ij}z_j+h_i\epsilon_i, \\\\\, \epsilon_i \sim TN[(\alpha_i-\sum_{j\neq i}c_{ij}z_j)/h_i, (\beta_i-\sum_{j\neq i}c_{ij}z_j)/h_i].$$

Denote the vectors of coefficients $c^{i}=(c_{i1}, \dots, c_{i, i-1}, c_{i, i+1}, \dots, c_{in})',  i=1, \dots, n$. From the conventional theory of the conditional multivariate normal distribution,  $c^i=-(T^{ii})^{-1}T^{i, <i} $ and $h_i^2=(T^{ii})^{-1},$
where $T^{ii}$ is the element in row $i$ and column $i$ of $T^{-1}$, and $T^{i, <i}$ is row $i$ of $T^{-1}$ with $T^{ii}$ deleted. 

Therefore, to sample ${\bm{a}}_{ij}$ from the posterior conditional probabilities, we compose a cycle of $K_{ij}$ Gibbs steps through the components of ${\bm{a}}_{ij}$. In the $k$th step of this cycle, $a_{ijk}$ is simulated from $a_{ijk}\mid{\bm{e}}^K, a_{ijq}(q\neq k), {\bm \beta}_i, m_i,  \Theta\}$, which is a univariate normal distribution truncated to one region. 

\subsubsection*{Sample the degree of polynomial $m_i$}
\begin{eqnarray*}
P(m_i \mid {\bm a}^K,\Theta)&\propto& P({\bm a}^K \mid \bmm, \Theta)P(m_i)=P(m_i) \int P({\bm a}^K \mid \bbeta, \bmm, \Theta)P(\bbeta \mid \bmm)~d\bbeta \nonumber\\
&\propto&P(m_i)  \frac{|A_{m_i}|^{\frac{1}{2}}}{|\sigma_0^2{\bf I}_{m_i+1}|^{\frac{1}{2}}}\nonumber\\
&&\times\exp\Big\{\frac{1}{2}\bb_{m_i}'A_{m_i}'\bb_{m_i}-\frac{1}{2}{\bm \mu}_0^{m_i+1'}(\sigma_0^2{\bm I}_{m_i+1})^{-1}{\bm \mu}_0^{m_i+1}\Big\}, \nonumber\\
\end{eqnarray*}
where $$A_{m_i}=\left\{\sum_{j=1}^{N_j}\bX_{ij}'\bC(\bt_{ij})^{-1}\bX_{ij}+(\sigma_0^2{\bm I}_{m_i+1})^{-1}\right\}^{-1},$$ $$\bb_{m_i}=\left\{\sum_{j=1}^{N_j}\bX_{ij}'\bC(\bt_{ij})^{-1}{\bm a}^K_{ij}+(\sigma_0^2{\bm I}_{m_i+1})^{-1}{\bm \mu}_0^{m_i+1}\right\}.$$

We only consider $(M+1)$ possible models, so we compute $r_{h}=P(h \mid {\bm a}^K,\Theta)$, $h=0, 1, \dots, M$. Let $w_{j}=r_{j}/\sum_{h=0}^Mr_{h}$, then we sample $m_i\sim \mathrm{Multinomial}(w_0, \dots, w_M)$. 

\subsubsection*{Posterior conditional distribution of $\bbeta_i$}
The posterior conditional distribution of $\bbeta_i$ is a multivariate normal distribution.
\begin{eqnarray*}
&&P({\bm \beta}_i\mid {\bm{a}}^K, \bmm, \Theta) \sim MVN(\bmu^{\beta}_i,\left\{\sum_{j=1}^{N_j}\bX_{ij}'\bC(\bt_{ij})^{-1}\bX_{ij}+(\sigma_0^2{\bm I}_{m_i+1})^{-1}\right\}^{-1}),\nonumber\\
\end{eqnarray*}
where
\begin{eqnarray*}
\bmu^{\beta}_i&=&\left\{\sum_{j=1}^{N_j}\bX_{ij}'\bC(\bt_{ij})^{-1}\bX_{ij}+(\sigma_0^2{\bm I}_{m_i+1})^{-1}\right\}^{-1}\nonumber\\
&&\times \left\{\sum_{j=1}^{N_j}\bX_{ij}'\bC(\bt_{ij})^{-1}{\bm a}^K_{ij}+(\sigma_0^2{\bm I}_{m_i+1})^{-1}{\bm \mu}_0^{m_i+1}\right\}. \nonumber\\
\end{eqnarray*}

\subsubsection*{Hybrid Monte Carlo algorithm to sample $\Theta$}
We apply Hybrid Monte Carlo method to sample $\Theta$.
Suppose we want to sample from the canonical distribution for a set of variables, $\Theta=\{\theta_1, \theta_2, \dots, \theta_n\}$, with respect to the potential energy function $E(\Theta)$. Assuming that $E(\Theta)$ is differentiable with respect to $\theta_i$, this canonical distribution is
\begin{equation*}
P(\Theta)=\frac{1}{Z_E}\exp\{-E(\Theta)\}.
\end{equation*}

We then introduce another set of variables, $\bw=\{w_1, w_2, \dots, w_n\}$, one $w_i$ for each $\theta_i$, with a kinetic energy function, $P(\bw)=\frac{1}{Z_K}\exp\{-K(w)\}=(2\pi)^{-n/2}\exp(-\frac{1}{2}\sum_iw_i^2)$. The total energy function, known as Hamiltonian, is $H(\Theta,\bw)=E(\Theta)+K(\bw)=E(\Theta)+\frac{1}{2}\sum_iw_i^2$. The canonical distribution defined by this energy function is $P(\Theta,\bw)=\frac{1}{Z_H}\exp\{-H(\Theta,\bw)\}=P(\Theta)P(\bw)$.

The unknown parameters $(\theta_1, \theta_2, \dots, \theta_n)$ can be
sampled from the marginal distribution for $\Theta$ by ignoring the
values we obtained for $\bw$. In practice, the differentiation
equations that describe the Hamiltonian equilibrium through time need to be discretized. In our case, to obtain the posterior samples of our unknown parameters, we use leapfrog discretization. In order to perform a leapfrog discretization, the derivative of the log of the posterior probability with respect to hyper-parameters is needed. 

A single leapfrog iteration calculates approximations $\hat{\theta}$ and $\hat{w}$ at time $\tau+\epsilon$ from $\hat{\theta}$ and $\hat{w}$ at time $\tau$ as follows:
$$\hat{w}_i(\tau+\frac{\epsilon}{2})=\hat{w}_i(\tau)-\frac{\epsilon}{2}\frac{\partial E}{\partial \theta_i}\{\hat{\theta}_i(\tau)\}$$
$$\hat{\theta}_i(\tau+\epsilon)=\hat{\theta}_i(\tau)+\epsilon\hat{w}_i(\tau+\frac{\epsilon}{2})$$
$$\hat{w}_i(\tau+\epsilon)=\hat{w}_i(\tau+\frac{\epsilon}{2})-\frac{\epsilon}{2}\frac{\partial E}{\partial \theta_i}\{\hat{\theta}_i(\tau+\epsilon)\}.$$

\begin{center}
\fbox{\fbox{\parbox{16cm}{{\bf{The hybrid Monte Carlo algorithm}}

Given values for the magnitude of the leapfrog stepsize, $\epsilon_0$, and the number of leapfrog steps, $L$, the dynamical transitions of the hybrid Monte Carlo algorithm operate as follows:
\begin{itemize}
\item{Randomly choose a direction, $\lambda$, for the trajectory with
    the two values $\lambda=+1$ representing a forward trajectory, and
    $\lambda=-1$ representing a backward trajectory, both being equally likely.}
\item{Starting from the current state, $(\theta,w)=(\hat{\theta}(0), \hat{w}(0))$, perform L leapfrog steps with a stepsize of $\epsilon=\lambda\epsilon_0$, resulting in the state $(\hat{\theta}(\epsilon L), \hat{w}(\epsilon L))=(\theta^*, w^*)$.}
\item{Regard $(\theta^*, w^*)$ as a candidate for the next state, as in the Metropolis algorithm: accepting it with probability $A\{(\theta,w), (\theta^*, w^*)\}=\min\big\{1, \exp\{-(H(\theta^*, w^*)-H(\theta, w))\}\big\}$, otherwise letting the new state be the same as the old one.}
\end{itemize}
}}}
\end{center}

For our problem, 
\begin{eqnarray*}
E(\Theta)&=&\frac{1}{2}\sum_{i=1}^2\sum_{j=1}^{N_i}\left\{({\bm{a}}^K_{ij}-\bX_{ij}{\bm \beta}_i)'\bC(\bt_{ij})^{-1}({\bm{a}}^K_{ij}-\bX_{ij}{\bm \beta}_i)+log(|\bC(\bt_{ij})|)\right\}\nonumber\\
&&-logP(\Theta),\nonumber\\
\end{eqnarray*}

\begin{eqnarray*}
\frac{\partial E(\Theta)}{\partial \Theta}
&=&-\frac{1}{2}\sum_{i=1}^2\sum_{j=1}^{N_i}({\bm{a}}^K_{ij}-\bX_{ij}{\bm \beta}_i)'\bC(\bt_{ij})^{-1}\frac{\partial \bC(\bt_{ij})}{\partial \Theta}\bC(\bt_{ij})^{-1}({\bm{a}}^K_{ij}-\bX_{ij}{\bm \beta}_i) \nonumber\\
&&+\frac{1}{2}\sum_{i=1}^2\sum_{j=1}^{N_i}tr\left\{\bC(\bt_{ij})^{-1}\frac{\partial \bC(\bt_{ij})}{\partial \Theta})\right\} -\frac{\partial logP(\Theta)}{\partial \Theta}.\nonumber \\
\end{eqnarray*}

Since $C_{uv}=C(t_u, t_v)=\theta_1^2\exp\left\{-r^2\sin^2(\frac{\pi(t_u-t_v)}{\theta_2})\right\}+\delta_{uv}J^2$,
and we have the formula $\frac{\partial C^{-1}}{\partial \theta}=-C^{-1}\frac{\partial C}{\partial \theta}C^{-1}$, therefore,

\begin{equation*}
C'_{\theta_1}(t_i, t_j)=\frac{\partial C}{\partial \theta_1}=2\theta_1\exp\left\{-r^2\sin^2(\frac{\pi(t_u-t_v)}{\theta_2})\right\},
\end{equation*}

\begin{equation*}
C'_{r}(t_i, t_j)=\frac{\partial C}{\partial r}=\theta_1^2\exp\left\{-r^2\sin^2(\frac{\pi(t_u-t_v)}{\theta_2})\right\}(-2r)\sin^2\left\{\frac{\pi(t_u-t_v)}{\theta_2}\right\},
\end{equation*}

\begin{eqnarray*}
&&C'_{r}(t_i, t_j)=\frac{\partial C}{\partial \theta_2}\nonumber\\
&=&2r^2\theta_1^2\exp\left\{-r^2\sin^2(\frac{\pi(t_u-t_v)}{\theta_2})\right\}sin\left\{\frac{\pi(t_u-t_v)}{\theta_2}\right\}cos\left\{\frac{\pi(t_u-t_v)}{\theta_2}\right\}\left\{\frac{\pi(t_u-t_v)}{\theta_2^2}\right\}.
\end{eqnarray*}

\clearpage

\section*{Appendix B: Proofs of Lemmas and Main Theorem}
\subsection*{B.1: Proof of Lemma 1}
\begin{proof}
Clearly $d(f, g)\geq 0$ and $d(f, g)=d(g,f)$. Also, we can easily show $d(f, g)=0$ if and only if $f=g$ a.s.. To prove $d(\cdot, \cdot)$ induces a metric, the last condition we need to verify is triangle inequality $d(f, h)\leq d(f, g)+d(g, h)$ for each $f, g, h\in \mathcal{F}$. 

Define $Q_{fg}=\{\epsilon: P(\{t: |f(t)-g(t)|>\epsilon\})<\epsilon\}$, then $d(f, g)=\mathrm{inf} \ Q_{fg}.$ So we need to show
\begin{equation}
 \mathrm{inf} \ Q_{fh} \leq \mathrm{inf} \ Q_{fg}+\mathrm{inf} \ Q_{gh}.\label{lemma1}
 \end{equation}
  Assuming $\epsilon_1\in Q_{fg}$ and $\epsilon_2\in Q_{gh}$, then
\begin{eqnarray*}
P(\{t: |f(t)-h(t)|>\epsilon_1+\epsilon_2\}) &\leq& P(\{t: |f(t)-g(t)|>\epsilon_1\})\nonumber\\
&&+P(\{t: |g(t)-h(t)|>\epsilon_2\}) \nonumber\\
&\leq& \epsilon_1+\epsilon_2 \nonumber
\end{eqnarray*}
So if $\epsilon_1\in Q_{fg}$ and $\epsilon_2\in Q_{gh}$, then $\epsilon_1+\epsilon_2\in Q_{fh}$, which implies (\ref{lemma1}).

Lastly we show $f_n$ converges to $f$ in probability if and only if $\mathrm{lim}_{n\rightarrow\infty}d(f_n, f)=0$. First, assume that $\mathrm{lim}_{n\rightarrow\infty}d(f_n, f)=0$. Then for every $\epsilon>0$ there exists $N$ such that for all $n\geq N$, $d(f_n, f)\leq \epsilon$, which is equivalent to $P(\{t: |f_n-f|>\epsilon\})<\epsilon$. Hence, $f_n$ converges to $f$ in probability. Finally, assume that $f_n$ converges to $f$ in probability. Then for every $\epsilon>0$, $\mathrm{lim}_{n\rightarrow\infty}P(\{t: |f_n-f|>\epsilon\})=0$. So, for every $\epsilon>0$, there exists $N$ such that for all $n\geq N$, $d(f_n, f)\leq \epsilon$, which completes the proof.
\end{proof}

\subsection*{B.2: Proof of Lemma 2}
\begin{proof}
It follows easily from applying Taylor's expansion to $\log(\frac{a}{b})$ and $\log(\frac{1-a}{1-b})$.  
\end{proof}

\subsection*{B.3: Proof of Lemma 3}

\begin{proof}
Our mean function $\mu(t)=\beta_0+\beta_1t$ is continuously differentiable in $[0, T_E]$. For sufficiently large $M$, 
\begin{eqnarray*}
Pr(\mathrm{sup}_t|a(t)|>M)&\leq& Pr(\mathrm{sup}_t|a(t)-\mu(t)|>M-\mathrm{sup}_t|\mu(t)|) \nonumber\\
&\leq&Pr(\mathrm{sup}_t|a(t)-\mu(t)|>M/2) \nonumber
\end{eqnarray*}
Thus, without generality, we assume that the mean function is identically zero.

From the result of Theorem 5 in \cite{ghosal2006posterior}, there exist constants $A_w, c_w$, $d^w(\theta_1, r, \theta_2)$ such that
\begin{eqnarray*}
Pr\left\{\mathrm{sup}_{t\in [0, T_E]}|D^wa(t)|>M_n\mid \theta_1, r, \theta_2\right\}\leq A_we^{-c_wd^wn}, \nonumber
\end{eqnarray*}
where $w=0, 1$ and $c_w>0$. Through calculations (details not shown),  $d^0=1/C_0(0; \theta_1, r, \theta_2)=\frac{1}{\theta_1^2}$, $d^1=-1/C''_0(0; \theta_1, r, \theta_2)=-\frac{1}{2\pi^2\theta_1^2r^2/\theta_2^2}$. Furthermore, since we assume $\rho_1(\theta_1, r, \theta_2)$ and $\rho_2(\theta_1, r, \theta_2)$ are continuous on the compact set $B$, they are uniformly bounded. Thus, there exist universal constant $S_1, S_2, S_3$ and $S_4$ such that
\begin{eqnarray*}
0&<&S_1\leq \mathrm{sup}_{(\theta_1, r, \theta_2)\in B}|\rho_1|\leq S_2 \nonumber\\
0&<&S_3\leq \mathrm{sup}_{(\theta_1, r, \theta_2)\in B}|\rho_2|\leq S_4 \nonumber
\end{eqnarray*}
Consequently, 
\begin{eqnarray*}
\mathrm{sup}_{(\theta_1, r, \theta_2)\in B}Pr\left\{\mathrm{sup}_{t\in [0, T_E]}|a(t)|>M_n\mid \theta_1, r, \theta_2\right\}\leq Ae^{-d_1n} \nonumber\\
\mathrm{sup}_{(\theta_1, r, \theta_2)\in B}Pr\left\{\mathrm{sup}_{t\in [0, T_E]}|a'(t)|>M_n \mid\theta_1, r, \theta_2\right\}\leq Ae^{-d_2n}, \nonumber
\end{eqnarray*}
where $d_1=c_0/S_2$, $d_1=c_1/S_4$ and $A=\mathrm{max}(A_0, A_1)$.
\end{proof}

\subsection*{B.4: Proof of Theorem 1}
\begin{proof}
In trials as the lupus trial, disease relapses frequently. Whenever
that happens, an binary outcome is observed and time
recorded. Assuming the relapse time is random, it is reasonable to
assume that 
$K\to \infty$ when $n\to \infty$. 
On the other hand, the theorem will not be valid if the observational
time points are regularly spaced and all patients are observed at the
same time. For example,  the theorem will not hold if $K_1=\dots= K_n=K_0$ and we observe each $e_{jk}$ only at the points $T_E/k$, where $k=1, \dots, K_0$.

The first condition of the Consistency Theorem in \cite{choi2007posterior} we need to verify is prior positivity of neighborhoods. If the
prior satisfies this condition, the probability of every
Kullback-Leibler (KL) neighborhood of true density function is
positive. The density of $e_{t_k}$ with respect to the counting
measure on $\{0,1\}$, say $Q$, is given by
$f(e_{t_k})=p(t_k)^{e_{t_k}}(1-p(t_k))^{1-e_{t_k}}$. For simplicity,
the index $t_k$ is dropped here. So we have $f(e)=p^e(1-p)^{1-e}$. The
corresponding true density function is $f_0(e)=p_0^e(1-p_0)^{1-e}$ 
and $\int
f_0\log(f_0/f)~dQ=p_0\log(p_0/p)+(1-p_0)\log((1-p_0)/(1-p))$.

To show that the probability of every KL neighborhood of true density function is positive, we first show that $\Pi\{f: \int f_0\log(f_0/f)~dQ<\epsilon\}>0$ for all $\epsilon>0$, where $\Pi$ is the prior for $f$, or equivalently, $\Pi\{p: p_0\log(p_0/p)+(1-p_0)\log((1-p_0)/(1-p))<\epsilon\}>0$ for all $\epsilon>0$. Let $\Upsilon=\{p: ||p-p_0||_{\infty}<\frac{1}{2}\epsilon_1\}$, where $\epsilon_1=\mathrm{inf}\{\min(p_0, 1-p_0)): 0\leq t\leq T_E\}>0$ and $||p-p_0||_{\infty}=\sup_{0\leq t\leq T_E}(|p(t)-p_0(t)|)$. If $p\in \Upsilon$,  following Lemma \ref{th:inequality}, there exists a constant $L$ depending only on $\epsilon_1$ such that $p_0\log(p_0/p)+(1-p_0)\log((1-p_0)/(1-p))\leq L||p-p_0||^2_{\infty}$. Therefore, it suffices to show that $\Pi(p: ||p-p_0||_{\infty}<\epsilon)>0$ for every $\epsilon>0$. Since our link function $H$ is assumed to be bounded and Lipschitz continuous, it suffices to show that $\Pi(a: ||a-a_0||_{\infty}<\epsilon)>0$ for every $\epsilon>0$. 

The prior distribution of $a$ is $a\sim GP(\mu, C)$, where $||\mu||_{\infty}<A_2$ and $||\mu'||_{\infty}<A_3$ for some constants $A_2$ and $A_3$ under our assumptions. Without loss of generality we assume $\mu \equiv 0$. 
\citet{choi2004posterior} examined the general result on the uniform support for a Gaussian process prior with zero mean. It is easily extended to support $[0, T_E]$ and our mean assumption. Therefore the positivity of neighborhoods holds.

The second condition is the existence of tests. We construct a similar sieve as in \citet{choi2004posterior} and then construct a test for each element of the sieve. 
$$SI_K=\{p(\cdot): p(t)=H(a(t)), ||D^wa||_{\infty}<M_K\}.$$
Let $p_1$ be a continuous function on $[0, T_E]$. Let $h_k=1$ if $p_1(t_k)\geq p_0(t_k)$ and $-1$ otherwise. Let $r>0$, $m_K=K^{1/2}$ and $I_K$ be the indicators of set $U_1=\{\sum_{k=1}^Kh_k(e_0(t_k)-p_0(t_k)>2m_K\sqrt{K}\}$, where $e_0(t_k)\sim \mathrm{Bernoulli}(p_0(t_k))$. For all $p_1$ that satisfy
\begin{eqnarray}
\label{con3}
\sum_{k=1}^K|p_1(t_k)-p_0(t_k)|>rK,
\end{eqnarray}
by Bernstein's inequality, we have 
\begin{eqnarray}
\label{con1}
\mathrm{E}_{P_0}(I_K)&=&P_0\left\{\sum_{k=1}^Kh_k(e_0(t_k)-p_0(t_k)>2m_K\sqrt{K}\right\} \nonumber\\
&=&2\exp(-\frac{1}{2}\frac{4m_K^2K}{K+2m_K\sqrt{K}})\leq 2\exp(-2m_K^2).
\end{eqnarray}

Also, let us assume $e(t_k)\sim \mathrm{Bernoulli}(p(t_k))$.  For all sufficiently large $K$ such that $m_K/\sqrt{K}<4/r$ and all $p$ satisfying $||p-p_1||_{\infty}<r/4$, we have
\begin{eqnarray}
\label{con2}
\mathrm{E}_{P}(1-I_K)&=&P\left\{\sum_{k=1}^Kh_k(e(t_k)-p_0(t_k)\leq2m_K\sqrt{K}\right\} \nonumber\\
&=&P\Bigg\{\frac{1}{\sqrt{K}}\sum_{k=1}^Kh_k(e(t_k)-p(t_k))+\frac{1}{\sqrt{K}}\sum_{k=1}^Kh_k(p(t_k)-p_1(t_k))\nonumber\\
&&+\frac{1}{\sqrt{K}}\sum_{k=1}^Kh_k(p_1(t_k)-p_0(t_k))\leq 2m_n\Bigg\}\nonumber\\
&\leq&P\left\{\frac{1}{\sqrt{K}}\sum_{k=1}^Kh_k(e(t_k)-p(t_k))\leq \frac{r\sqrt{K}}{4}-r\sqrt{K}+2m_K\right\}\nonumber\\
&\leq& 2\exp(-\frac{1}{2}\frac{r^2}{1+r}K),
\end{eqnarray}
where the second to last inequality is by Bernstein's inequality. 

We showed how to construct consistent test functions when the inequality (\ref{con3}) holds. \cite{choi2007alternative} examined inequality (\ref{con3}) held with metric $d$ introduced in Lemma \ref{distance}. By (\ref{con1}), (\ref{con2}) and Lemma \ref{gpineq}, the second condition of Consistency Theorem in \cite{choi2007alternative} is verified.

So far the two conditions of Consistency Theorem in \cite{choi2007posterior} have been verified. We have showed that the posterior probability 
\begin{eqnarray}
\mathrm{sup}_{(\theta_1, r, \theta_2)\in B} \Pi\{S^2_{\epsilon}\mid \eb,  \theta_1, r, \theta_2\}\rightarrow 0  \ \   \ [P_0^{\infty}],
\label{eq2}
\end{eqnarray}
where $S^2_{\epsilon}=\left\{p: \int |p(t)-p_0(t)|dt>\epsilon\right\}$ 
and $\eb=(e_{t_1}, \dots, e_{t_K})'$.

Finally, let us consider the goal of our theorem. By Fubini's Theorem, 
\begin{eqnarray}
\Pi\left\{S^2_{\epsilon}\mid \eb\right\}&=&\int_B \Pi\{S^2_{\epsilon}\mid \eb, \theta_1, r, \theta_2\}~d\Pi\left\{(\theta_1, r, \theta_2)\mid\eb\right\} \nonumber\\
&=&\int_{B_3}\int_{B_2}\int_{B_1} \Pi\{S^2_{\epsilon}\mid\eb, \theta_1, r, \theta_2\}~d\Pi(\theta_1\mid\eb)~d\Pi(r\mid\eb) ~d\Pi(\theta_2\mid\eb).  \nonumber
\end{eqnarray}
Since the supermum of the conditional probability in (\ref{eq2}) converges to 0 in $P_0^K$ probability, the marginal posterior probability converges to 0 in $P_0^K$ probability regardless of the asymptotic distribution of $\Pi(\theta_1\mid\eb)$,  $\Pi(r\mid\eb)$ and  $\Pi(\theta_2\mid\eb)$. This is formalized as follows:
\begin{eqnarray}
\Pi\{S^2_{\epsilon}\mid \eb\}&=&\int_{B_3}\int_{B_2}\int_{B_1} \Pi\{S^2_{\epsilon}\mid\eb, \theta_1, r, \theta_2\}~d\Pi(\theta_1\mid\eb)~d\Pi(r\mid\eb) ~d\Pi(\theta_2\mid\eb)  \nonumber\\
&\leq&\mathrm{sup}_{(\theta_1, r, \theta_2)\in B} \Pi\{S^2_{\epsilon}\mid\eb,  \theta_1, r, \theta_2\}\int_{\theta_1\in B_1}~d\Pi(\theta_1\mid\eb) \nonumber\\
&&\times \int_{r\in B_2}d\Pi(r\mid\eb)\int_{\theta_2\in B_3}d\Pi(\theta_2\mid\eb) \ \ \rightarrow 0    \ \   \ [P_0^{\infty}] \nonumber
\end{eqnarray}
So, (\ref{consistency2}) is proved. From Lemma \ref{distance},
(\ref{consistency1}) is also proved. 
\end{proof}

\clearpage
\section*{Appendix C}
\noindent {\bf Theorem (Choi and Schervish, 2007)} 
{\it Let $\{Z_i\}_{i=1}^{\infty}$ be independently distributed with densities $\{f_i(\cdot; \theta)\}_{i=1}^{\infty}$, with respect to a common $\sigma$-finite measure, where the parameter $\theta$ belongs to an abstract measurable space $\Theta$. The densities $f_i(\cdot; \theta)$ are assumed to be jointly measurable. Let $\theta_0\in \Theta$ and let $P_{\theta_0}$ stand for the joint distribution of $\{Z_i\}_{i=1}^{\infty}$ when $\theta_0$ is the true value of $\theta$. Let $\{U_n\}_{i=1}^{\infty}$ be a sequence of subsets of $\Theta$. Let $\theta$ have prior $\Pi$ on $\Theta$. Define
$$\Lambda_i(\theta_0, \theta)=\log \frac{f_i(Z_i; \theta_0)}{f_i(Z_i; \theta)},$$
$$K_i(\theta_0, \theta)=\mathrm{E}_{\theta_0}(\Lambda_i(\theta_0, \theta)),$$
$$V_i(\theta_0, \theta)=\mathrm{Var}_{\theta_0}(\Lambda_i(\theta_0, \theta)).$$

(A1) Prior positivity of neighborhoods.

Suppose that thee exists a set $B$ with $\Pi(B)>0$ such that
\begin{enumerate}
\item $\sum_{i=1}^{\infty}\frac{V_i(\theta_0, \theta)}{i^2}<\infty$, $\forall \theta\in B$,
\item For all $\epsilon>0$, $\Pi(B \cap \{\theta: K_i(\theta_0, \theta)<\epsilon$ for all $i$\}$)>0.$ 
\end{enumerate}

(A2) Existence of tests

Suppose that there exist test functions $\{\Phi_n\}_{i=1}^{\infty}$, sets $\{\Theta_n\}_{i=1}^{\infty}$ and constants $C_1, C_2$, $c_1, c_2>0$ such that
\begin{enumerate}
\item $\sum_{i=1}^{\infty}\mathrm{E}_{\theta_0}\Phi_n<\infty$,
\item $\sup_{\theta \in U_n^C\cap \Theta_n} \mathrm{E}_{\theta}(1-\Phi_n)\leq C_1e^{-c_1n}$,
\item $\Pi(\Theta_n^C)\leq C_2e^{-c_2n}$.
\end{enumerate}

Then
$$\Pi(\theta\in U_n^C|Z_1, \dots, Z_n)\rightarrow 0  \ \ \ a.s. \ \ [P_{\theta_0}].$$

}

\clearpage
\bibliographystyle{apalike}
\bibliography{Gaussian}

\end{document}